\documentclass[aip,cha,amsmath,amssymb,graphicx,reprint]{revtex4-2}

\usepackage[utf8]{inputenc}
\usepackage[T1]{fontenc}
\usepackage{mathptmx}
\usepackage{etoolbox}
\usepackage{amsmath}

\usepackage{graphicx}
\usepackage{caption}
\usepackage{subcaption}

\usepackage{mathtools}
\usepackage{amsthm}

\def\I{\mathrm I}

\usepackage{graphicx,amssymb,amsmath}

\usepackage{bm}

\newcommand{\pmat}[1]{\begin{pmatrix} #1 \end{pmatrix}}
\newcommand{\E}{\mathbb{E}}
\newcommand{\tr}{\mathsf{T}}

\usepackage{comment}

\usepackage{lineno}

\begin{document}


\title{A stochastic approximation for the finite-size Kuramoto-Sakaguchi model} 



\author{Wenqi Yue}
\email[]{wyue8667@uni.sydney.edu.au}
\affiliation{School of Mathematics and Statistics, The University of Sydney, NSW, Australia}

\author{Georg A. Gottwald}
\email[]{georg.gottwald@sydney.edu.au}
\affiliation{School of Mathematics and Statistics, The University of Sydney, NSW, Australia}


\date{\today}

\begin{abstract}
We perform a stochastic model reduction of the Kuramoto-Sakaguchi model for finitely many coupled phase oscillators with phase frustration. Whereas in the thermodynamic limit coupled oscillators exhibit stationary states and a constant order parameter, finite-size networks exhibit persistent temporal fluctuations of the order parameter. These fluctuations are caused by the interaction of the synchronized oscillators with the non-entrained oscillators. We present numerical results suggesting that the collective effect of the non-entrained oscillators on the synchronized cluster can be approximated by a Gaussian process. This allows for an effective closed evolution equation for the synchronized oscillators driven by a Gaussian process which we approximate by a two-dimensional Ornstein-Uhlenbeck process. Our reduction reproduces the stochastic fluctuations of the order parameter and leads to a simple stochastic differential equation for the order parameter.
\end{abstract}


\maketitle


\section{Introduction}
Ever since Huygen's observation of two pendulum clocks, mounted on the same wall a short distance apart, ending up swinging in anti-phase, the phenomenon of collective behaviour and synchronisation of weakly coupled oscillators has fascinated scientists. Synchronization has been observed in a diverse range of natural and engineered  systems \cite{Kuramoto84,PikovskyEtAl01,Strogatz}, including in pace-maker cells of circadian rhythms \cite{YamaguchiETAl03}, networks of neurons \cite{BhowmikShanahan12}, in chemical oscillators \cite{KissEtAl02,TaylorEtAl09} and in power grid systems \cite{FilatrellaEtAl08}. 

The celebrated Kuramoto model of sinusoidally coupled phase oscillators has served as a rich model to study synchronisation \cite{Kuramoto84, Strogatz00, PikovskyEtAl01, AcebronEtAl05, OsipovEtAl07, ArenasEtAl08, DorflerBullo14, RodriguesEtAl16}. The Kuramoto model was extended to the Kuramoto-Sakaguchi model \cite{SakaguchiKuramoto86} to include the effect of time-delayed or phase frustrated coupling which was observed in numerous real-world contexts  \cite{CrookEtAl97}, including in arrays of Josephson junctions \cite{WiesenfeldEtAl96,BarbaraEtAl99,FilatrellaEtAl00}, in power grids \cite{NishikawaMotter15} and in seismology \cite{Scholz10,VasudevanEtAl15}. Apart from the usual transition from an incoherent state at low coupling strength to synchronisation upon increasing the coupling strength, the Kuramoto-Sakaguchi model exhibits a plethora of dynamical behaviours including bi-stability of incoherence and partial synchronisation, transition from coherence to incoherence with increasing coupling strength \cite{Omelchenko12,Omelchenko13}, chaotic dynamics \cite{BickEtAl18} as well as chimera states \cite{AbramsEtAl08,Laing09,MartensEtAl16}. 

The possible high dimensionality of networks of oscillators inhibits an understanding of the underlying dynamic mechanisms which give rise to this rich behaviour. Scientists have therefore looked at model reductions of the Kuramoto model and of the Kuramoto-Sakaguchi model. Most methods are restricted to the thermodynamic limit of infinitely many oscillators. In this limit Kuramoto and Sakaguchi established a mean-field theory which determines the order parameter and the non-zero rotation frequency of the synchronised cluster via a self-consistency relationship \cite{SakaguchiKuramoto86}. Similarly, the celebrated Ott-Antonson ansatz \cite{OttAntonson08} can be employed to obtain a deterministic evolution equation for the order parameter \cite{Omelchenko12,Omelchenko13}. Real world networks, however, are of finite size, and in finite-size systems the onset of synchronisation occurs typically not at the critical coupling strength predicted by the thermodynamic limit. While in the thermodynamic limit, the order parameter asymptotically in time approaches a constant value, in finite-size networks the order parameter exhibits persistent temporal fluctuations. Furthermore, finite-size networks exhibit a singularity in the variance of the order parameter at the onset of the transition to synchronisation with a well-defined finite size scaling \cite{Daido90,HongEtAl15,HongEtAl16}. To overcome the restriction of the thermodynamic limit and to tackle the case of finite-size networks, a collective coordinate approach \cite{Gottwald15,HancockGottwald18,SmithGottwald20} was applied to study the Kuramoto-Sakaguchi model for finitely many oscillators and derive an evolution equation for the order parameter \cite{YueEtAl20}. It was established that the order parameter and the dynamics of the synchronised cluster is markedly influenced by the dynamics of the non-entrained rogue oscillators. Describing the effect of the rogue oscillators by their average, a deterministic reduced evolution equation for the entrained oscillators was derived. This allowed for the estimation of the onset of synchronisation and the determination of the average behaviour of the order parameter. 

Whereas our previous work captured the averaged effect of the rogue oscillators on the synchronised oscillators, in this work we set out to quantitatively describe the fluctuations around this average behaviour and capture the effective stochastic dynamics of the synchronised phase-oscillators and the order parameter. We show how the thermodynamic limit is approached for increasing number of oscillators. 
We first present numerical results suggesting that the fluctuations constitute a Gaussian process.  
In a second step we approximate the Gaussian process by an easily computable Markovian Ornstein-Uhlenbeck process. This allows for a stochastic model reduction of the deterministic Kuramoto-Sakaguchi equation. In particular, we will propose a closed set of equations for the entrained synchronized oscillators where the effect of the non-entrained rogue oscillators is modelled by coloured noise, the variance of which decreases as the number of oscillators increases. We show numerically that the statistics of the order parameter and of the mean phase is well recovered by the reduced stochastic equation for the entrained oscillators. The stochastic approximation depends only, as we show numerically, on the number of the non-entrained rogue oscillators $N_r$. For $N_r \to \infty$ the variance of the noise goes to zero and the deterministic thermodynamic limit is approached. On the other extreme, the validity of the stochastic approximation requires a sufficiently large number of rogue oscillators for the validity of the central limit theorem. This implies, as we will show, that to observe an effective noise requires different network sizes for different values of the coupling strength: for large values of the coupling strength with a high degree of synchronization there are proportionally fewer rogue oscillators than for smaller values of the coupling strength, requiring a larger total number of oscillators $N$ to allow for sufficiently large numbers $N_r$ of rogue oscillators. 
We further use the stochastic model reduction to formulate a stochastic differential equation (SDE) for the order parameter. SDEs driven by Brownian motion for the order parameter were recently postulated and determined using a data-driven approach \cite{SnyderEtAl21}. We show here that the SDEs are in fact driven by coloured noise. 

There are two dynamic mechanisms capable of generating effective stochastic behaviour in deterministic systems: multiscale dynamics and weak coupling \cite{GivonEtAl04}. Both mechanisms generate stochasticity via some (functional) central limit theorems. In multiscale dynamics, slow variables experience on a long diffusive time scale the integrated effect of fast variables. If the fast dynamics is sufficiently chaotic, the integrated effect may lead to Brownian motion \cite{MelbourneStuart11,GottwaldMelbourne13,KellyMelbourne16,KorepanovEtAl22,ChevyrevEtAl22}. In weakly coupled systems in which resolved variables are weakly coupled to a large number of unresolved degrees of freedom, the sum over the many unresolved variables with uncorrelated initial conditions leads to a Gaussian process \cite{Kahane,KupfermanEtAl02,GivonEtAl04}. We establish here that the dynamic mechanism responsible for the effective stochastic dynamics of the collective behaviour of large but finite Kuramoto-Sakaguchi models is via the route of weak coupling.

The paper is organised as follows. In Section~\ref{sec:model} we introduce the Kuramoto-Sakaguchi model. Section~\ref{sec:mft} reviews the mean-field theory for the Kuramoto-Sakaguchi model. Section~\ref{sec:num} presents numerical simulations of the Kuramoto-Sakaguchi model illustrating its stochastic order parameter fluctuations. We develop our stochastic model reduction in Section~\ref{sec:sde}, in which we formulate an SDE for the entrained oscillators. We further use the stochastic model equations to formulate an SDE for the order parameter. In Section~\ref{sec:comp} we present numerical results of the reduced stochastic model and its ability to capture observed statistical properties of the full Kuramoto-Sakaguchi model. We conclude in Section~\ref{sec:disc} with a discussion.


\section{The Kuramoto-Sakaguchi model}
\label{sec:model}
The Kuramoto-Sakaguchi model \cite{Kuramoto84,SakaguchiKuramoto86,Strogatz00} 
\begin{align} 
\label{eq:ks}
\dot{\theta}_i = \omega_i + \frac{K}{N}\sum_{j=1}^N \sin(\theta_j-\theta_i-\lambda),\qquad i = 1,\dots N.
\end{align}
is a classic paradigmatic model describing the dynamics of $N$ sinusoidally globally coupled oscillators with phases $\theta_i$ under global phase frustration $\lambda$ with coupling strength $K$. Each oscillator is equipped with an intrinsic frequency $\omega_i$ which is drawn from a specified distribution $g(\omega)$. 

The Kuramoto-Sakaguchi model (\ref{eq:ks}) displays a transition to synchronisation as the coupling strength $K$ increases. For low values of $K$, the system is in an incoherent state in which each oscillator evolves approximately with their own intrinsic frequency. As the coupling strength $K$ increases, some of the oscillators become synchronised, oscillating with a common frequency and with their phases staying close to one another. When $K$ increases further, more and more oscillators become synchronised, until eventually global synchronisation occurs. The non-zero phase frustration $\lambda \neq 0$ induces a collective rotation of the synchronised cluster with a non-zero frequency $\Omega$ in the rest frame, as opposed to the Kuramoto model which supports stationary synchronised clusters. We remark that for $\lambda=0$ and uniform intrinsic frequency distributions a first-order phase transition occurs \cite{Pazo05}.

The collective behaviour can be described by the mean-field variables $r$ and $\psi$ with
\begin{align} 
	r(t)e^{i{\psi}(t)} = \frac{1}{N}\sum_{j=1}^{N}e^{i\theta_j(t)}.
\label{eq:r0}
\end{align}
The degree of synchronisation is quantified by the order parameter $\bar r$ with
\begin{align*} 
	\bar{r}=\lim_{T\rightarrow\infty}\frac{1}{T}\int_{0}^{T}r(t)dt.
\end{align*}
Perfect phase synchronisation with $\theta_1=\theta_2=\cdots=\theta_N$ implies $\bar r=1$ and $\bar r\gtrsim0$ indicates that phases are spread out with $\bar r \sim 1/\sqrt{N}$ indicating incoherence. 


\subsection{Classical mean-field theory}
\label{sec:mft}
Sakaguchi and Kuramoto \cite{SakaguchiKuramoto86} developed a mean-field theory for the mean frequency $\Omega$ of the synchronzied cluster and the order parameter $r$ which we present here for completeness. We follow here their exposition to obtain self-consistency relations for the mean-field variables $r$ and $\Omega$ as well as expressions for the stationary density function. The stationary density will be used later to determine the average effect of the rogue oscillators onto the synchronized oscillators. The assumption in the Kuramoto-Sakaguchi model \eqref{eq:ks} that each oscillator is connected to all other oscillators is crucial for the following derivation. Moving into the frame of reference rotating with the cluster mean frequency $\Omega=\Omega(K)$ and setting the mean-field phase variable $\psi=0$, the Kuramoto-Sakaguchi model (\ref{eq:ks}) is expressed as 
\begin{align}
\label{eq:ksmf}
\dot{\theta}_i(t)=v(\theta_i;\omega_i),
\end{align}
with frequency
\begin{align} 
\label{v_mean_field}
v(\theta_i;\omega_i)=\omega_i-\Omega-Kr\sin(\theta_i+\lambda).
\end{align}
Each oscillator $\theta_i$ only couples to the other oscillators via the mean-field in the form of $r$ and $\Omega$.

From (\ref{eq:ksmf}) one can readily identify the oscillators which form the synchronised cluster and the rogue oscillators which are not entrained: The former ones have frequencies $|\omega_i-\Omega|\leq Kr$ for which (\ref{eq:ksmf}) has stationary solutions with
\begin{align*} 
\theta_i=\arcsin\left(\frac{\omega_i-\Omega}{Kr}\right)-\lambda.
\end{align*}
The synchronized oscillators rotate with the collective mean frequency $\Omega$. In contrast, the non-entrained rogue oscillators have frequencies $|\omega_i-\Omega|>Kr$ and satisfy the Adler equation (\ref{eq:ksmf}).  

In the thermodynamic limit $N \rightarrow \infty$, the phases can be described by a probability density function $\rho(\theta,t;\omega)$ satisfying the continuity equation
\begin{align} 
\label{eq:rho_t}
\frac{\partial \rho}{\partial t}+\frac{\partial}{\partial \theta}(\rho v)=0.
\end{align}
For stationary states with $\frac{\partial \rho}{\partial t} = 0$ there are two kinds of stationary density solutions, depending on the intrinsic frequency $\omega$. Entrained oscillators with $|\omega_i-\Omega|\leq Kr$ which form the synchronised cluster are captured by the stationary probability density function
\begin{align} 
\label{eq:rho_entrained}
\rho(\theta;\omega)=\delta(\theta-\arcsin\left(\frac{\omega-\Omega}{Kr}\right)+\lambda). 
\end{align}
The non-entrained rogue oscillators with $|\omega_i-\Omega| >Kr$ are captured by the stationary probability density function
\begin{equation} 
\label{eq:rho_rogue}
\rho(\theta;\omega)=\frac{C(\omega)}{v(\theta;\omega)}=\frac{C(\omega)}{\omega-\Omega-Kr\sin(\theta+\lambda)},
\end{equation}
with normalization constant $C(\omega)$. In the thermodynamic limit the order parameter 
\begin{align*} 
r=\int_{-\infty}^{\infty} \int_{0}^{2\pi}e^{i \theta}\rho(\theta,t;\omega) g(\omega) d\theta d\omega,
\end{align*}
for the mean-field Kuramoto-Sakaguchi equation (\ref{eq:ksmf}) is then given by separating the integration over the frequencies into the entrained and non-entrained ranges, with their respective stationary densities \eqref{eq:rho_entrained} and \eqref{eq:rho_rogue}. We obtain 
\begin{align}
re^{i \lambda}&= \; \,\,
\displaystyle\int\displaylimits_{\mathclap{|\omega-\Omega|\leq Kr}} \;\;\;\left(\sqrt{1-\frac{(\omega-\Omega)^2}{K^2r^2}}+ i\frac{\omega-\Omega}{Kr}\right) g(\omega)d \omega \nonumber\\
&+i\displaystyle\int\limits_{\mathclap{|\omega-\Omega|>Kr}}\;\;\; \left(\frac{\omega-\Omega}{Kr}-\frac{\omega-\Omega}{Kr}\sqrt{1-\frac{K^2r^2}{(\omega-\Omega)^2}}\right)g(\omega)d \omega. 
\label{eq:self_consistency_1}
\end{align}
Separating the real and imaginary parts of (\ref{eq:self_consistency_1}), we arrive at
\begin{align} 
r \cos\lambda&=\displaystyle\int\displaylimits_{\mathclap{|\omega-\Omega|\leq Kr}} \sqrt{1-\frac{(\omega-\Omega)^2}{K^2r^2}} g(\omega) d\omega
\label{eq:self_consistency_real}\\
r \sin \lambda&= \displaystyle\int\displaylimits_{\mathclap{|\omega-\Omega|\leq Kr}} \;\; \frac{\omega-\Omega}{Kr}g(\omega)d\omega \nonumber\\
&+\displaystyle\int\limits_{\mathclap{|\omega-\Omega|>Kr}}\;\;\; \left(\frac{\omega-\Omega}{Kr}-\frac{\omega-\Omega}{Kr}\sqrt{1-\frac{K^2r^2}{(\omega-\Omega)^2}}\right)g(\omega)d \omega.
\label{eq:self_consistency_imag}
\end{align}
These are self-consistency relations for the mean-field variables $r$ and $\Omega$ and can be numerically solved for any given intrinsic frequency distribution $g(\omega)$, coupling strength $K$ and phase frustration parameter $\lambda$. In the following we denote the solution of the order parameter of the self-consistency relation by $r_\infty$.

%


\section{Finite-size effects in the Kuramoto-Sakaguchi equation}
\label{sec:num}
We now present results of numerical simulations of the Kuramoto-Sakaguchi model (\ref{eq:ks}), illustrating finite-size effects. We employ a 4th order Runge-Kutta(RK4) method with time step $dt=0.05$. Initial conditions are chosen randomly from the interval $[0,2\pi]$; to eliminate transient behaviour we discard a transient period of $t_0 = 5\times 10^4$ time units. Statistics such as means and variances are computed from time series of length $t_{\textrm{max}} = 5\times 10^4$. In the following we fix the phase frustration parameter to $\lambda=\pi/4$ and consider a Gaussian distribution of the intrinsic frequencies with mean $0$ and variance $1$. To avoid sampling effects which may lead to small cluster nucleation \cite{PeterPikovsky18,FialkowskiEtAl23} we consider here equiprobable intervals between the $N$ intrinsic frequencies \footnote{For a given system size $N$,  the frequencies $\omega_i$ are generated deterministically by dividing $[0,1]$ into $N+1$ equiprobable intervals according to the distribution $g(\omega)$ and taking the $N$ end-points of the intervals (excluding the boundaries at $0$ and $1$) to find the corresponding frequencies $\omega_i$. i.e. the $\omega_i$'s are obtained by solving $\int_{-\infty}^{\omega_i} g(\omega) d\omega = i/(N+1), i = 1,2,\dots,N.$ When reporting $N$ in the legends of the figures, it refers to the number of chosen intervals.}. This allows us to compare the effect of different system sizes $N$ where for each system size $N$ the intrinsic frequencies are uniquely fixed, and eliminates variations due to particular draws of $\omega_i$ for a given system size $N$. For a detailed discussion on the effects of random sampling see, for example, \cite{HongEtAl15,PeterPikovsky18}. Unless specified otherwise we set $K=3$ which corresponds to a partially synchronized state of the system where both synchronized and non-synchronized oscillators occur.\\ 

\begin{figure}[]
\centering
\begin{subfigure}[]{\linewidth}
\centering
\includegraphics[width = 0.9\linewidth]{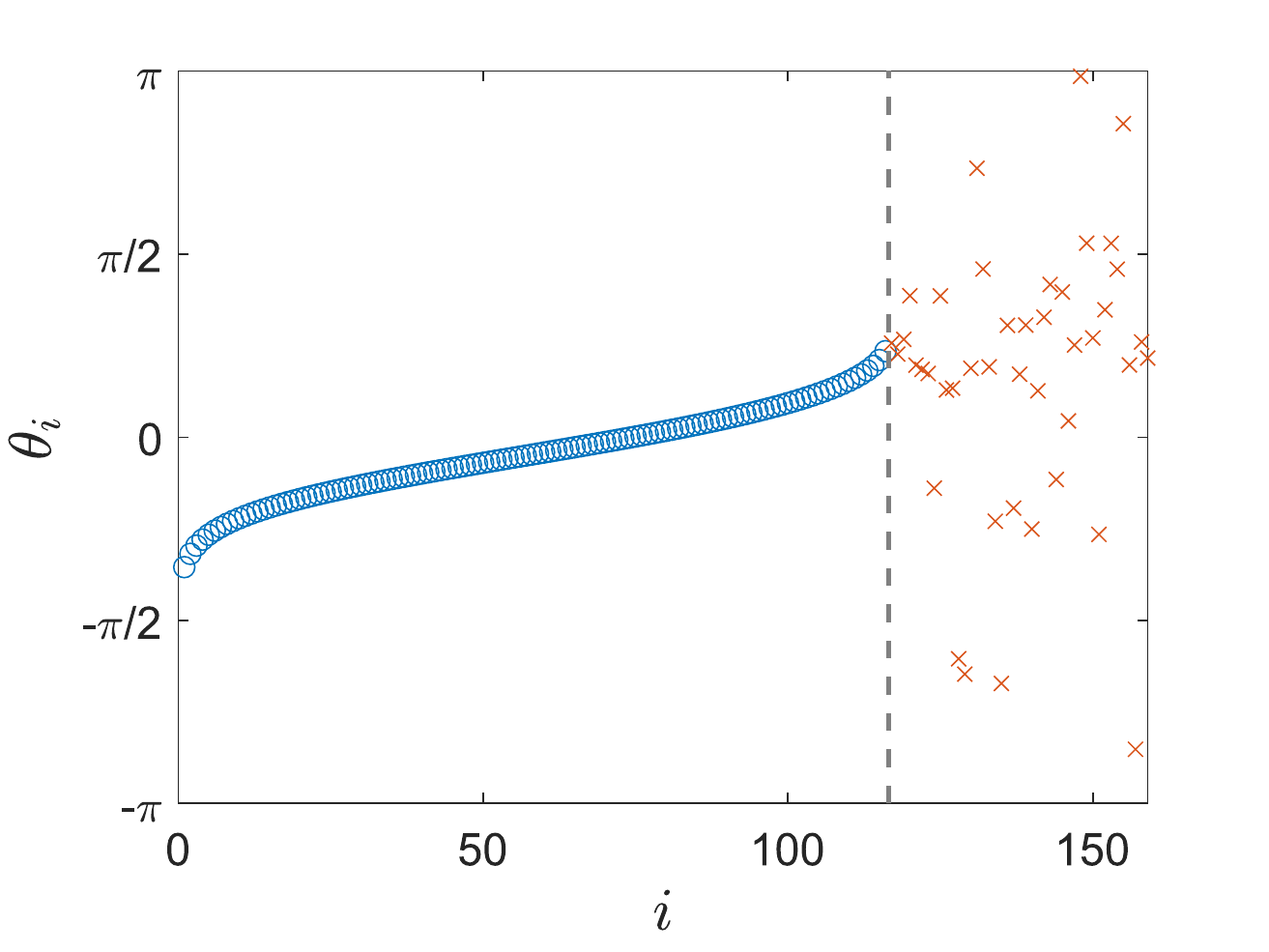}
\caption{}
\end{subfigure}
\hfill
\begin{subfigure}[]{\linewidth}
\centering
\includegraphics[width = 0.9\linewidth]{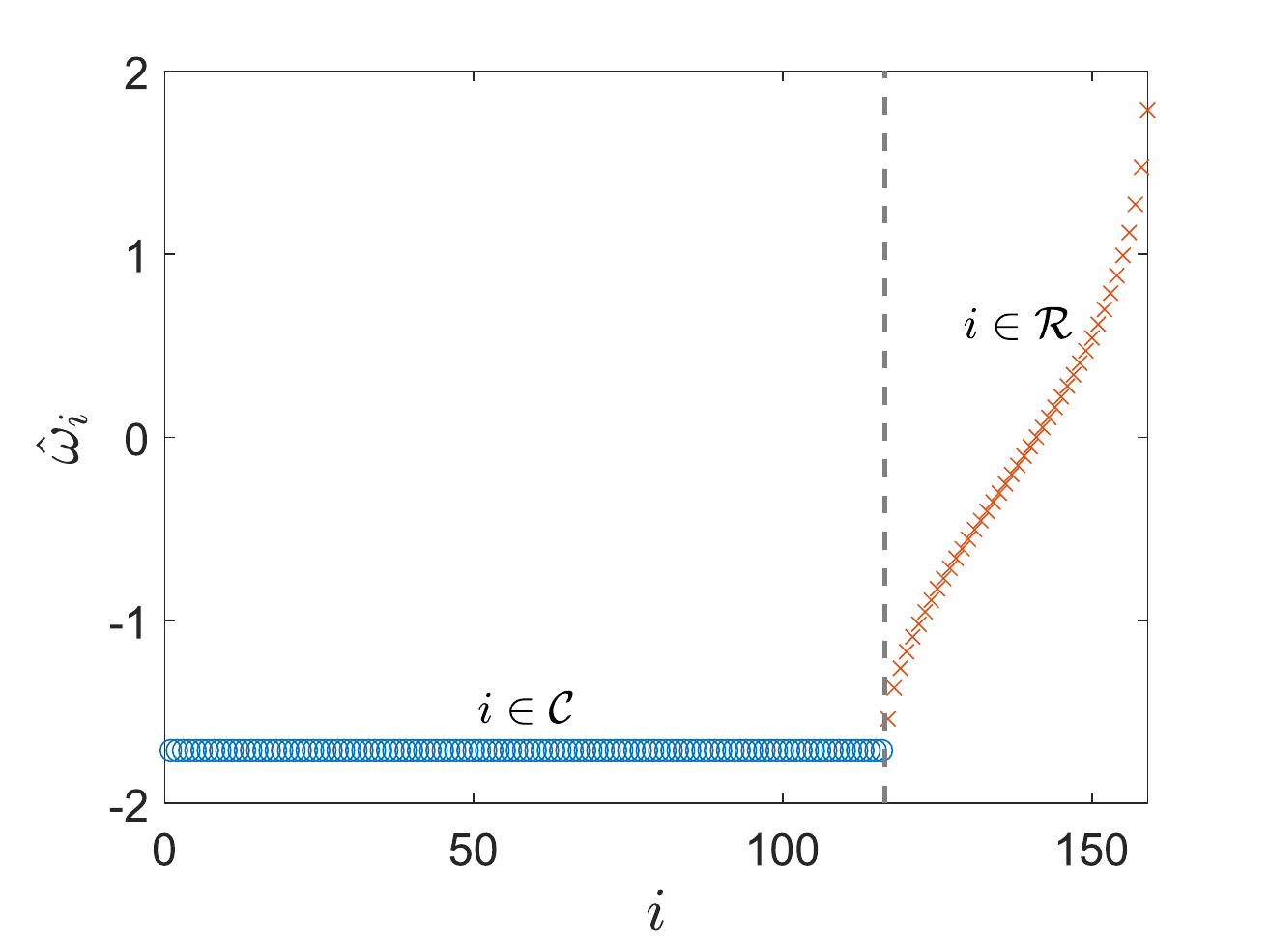}
\caption{}
\end{subfigure}
\caption{Snapshot of the phases $\theta_i$ (a) and effective frequencies $\hat\omega_i$ (b) for the Kuramoto-Sakaguchi model (\ref{eq:ks}) with $N = 160$ oscillators and phase frustration $\lambda=\frac{\pi}{4}$ at coupling strength $K=3$ with a Gaussian intrinsic frequency distribution with zero mean and unit variance. Circles (online blue) denote oscillators entrained in collective synchronised dynamics. Crosses (online red) denote the non-entrained rogue oscillators. We identify oscillators with indices $i\le 116$ as the synchronized oscillators with $N_c=116$ and oscillators with indices $i>116$ as the rogue oscillators with $N_r=43$.}
\label{fig:snapshot}
\end{figure}

Figure \ref{fig:snapshot}a shows a snapshot of the phases for $N=160$. The phases are labelled according to their intrinsic frequencies such that the oscillator with the most negative intrinsic frequency is assigned the label $i=1$ and the one with the largest frequency the label $i=N$. One can see clearly the synchronised cluster with frequencies close to the mean frequency $\Omega$, and the non-entrained rogue oscillators with more extreme intrinsic frequencies. Note that due to the non-zero phase frustration $\lambda$ the synchronised cluster is not centred around the oscillators closest to the mean of their respective intrinsic frequencies as is the case for the standard Kuramoto model with $\lambda=0$. To determine the synchronised cluster and its complement, the non-entrained rogue oscillators, we define the effective frequency of each oscillator,
\begin{align*} 
	\hat \omega_i = \left\langle\dot{\theta}_i(t)\right\rangle_t,
\end{align*}
where the angular brackets denote a temporal average. Those oscillators with approximately the same effective frequencies $\hat\omega_i$ are considered to form the synchronised cluster $\mathcal{C}$ of size $N_c$, and the remaining oscillators are considered to be within the set of non-synchronised rogue oscillators $\mathcal{R}$ with size $N_r$. Figure~\ref{fig:snapshot}b shows the effective frequencies. We estimate the number of rogue oscillators for $K=3$ and $N=160$ oscillators to be $N_r=43$. We identify oscillators with indices $i\le 116$ as the synchronized oscillators with $N_c=116$ and oscillators with indices $i>116$ as the rogue oscillators.\\

\begin{figure}[] 
\centering
	\includegraphics[width = 0.9\linewidth]{{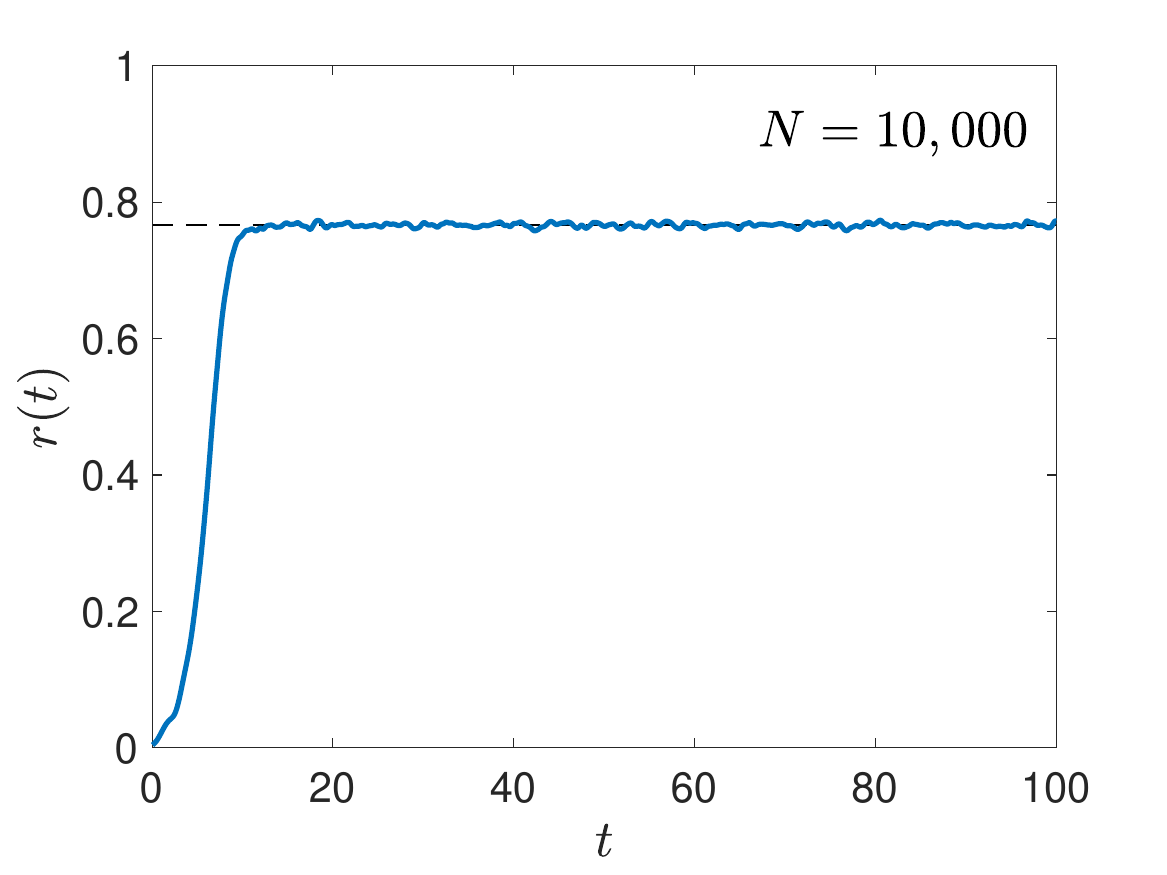}}
	\includegraphics[width = 0.9\linewidth]{{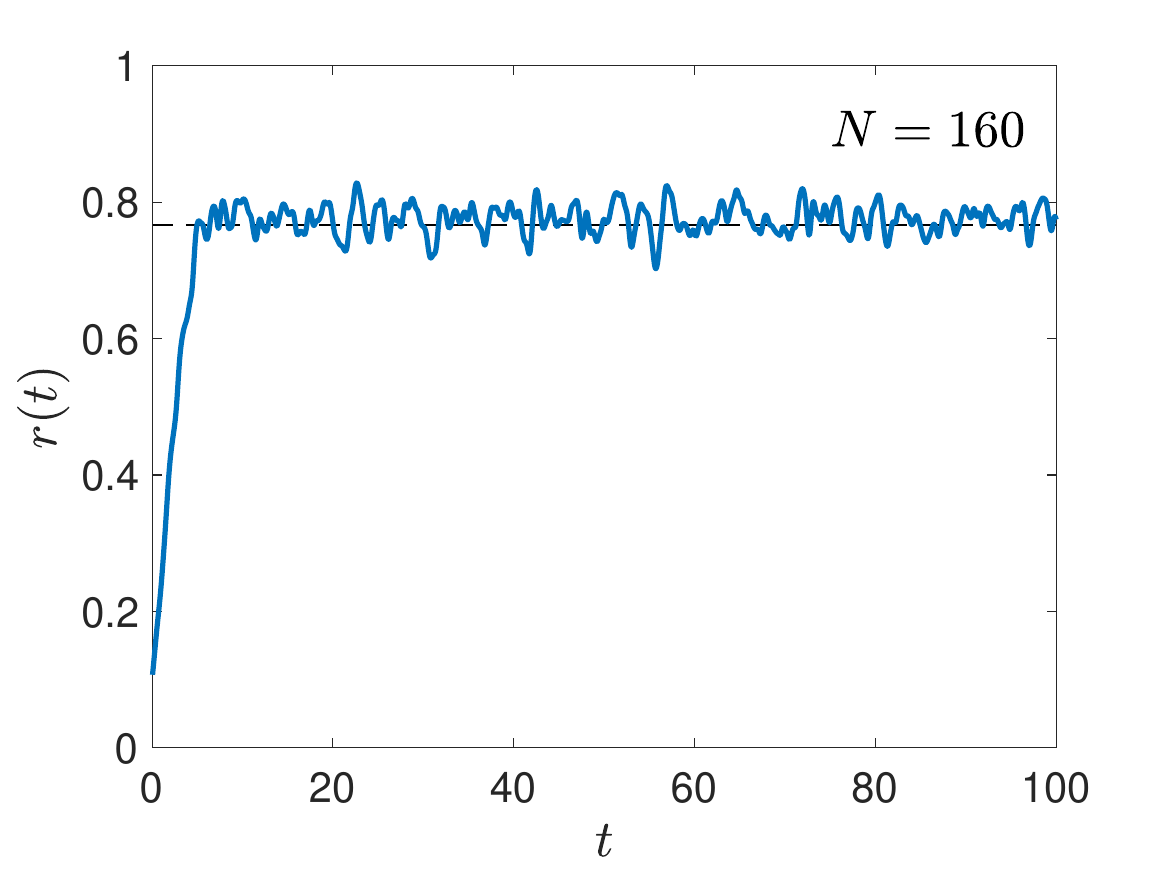}}
	\caption{Time series of the order parameter $r(t)$ from a single simulation of the Kuramoto-Sakaguchi model (\ref{eq:ks}) with $N=10,000$ and $N = 160$ oscillators, starting from a random initial condition. The dashed horizontal line indicates the stationary value $r_\infty=0.766$ of the corresponding thermodynamic limit, estimated by solving the self-consistency relations (\ref{eq:self_consistency_real})-(\ref{eq:self_consistency_imag}). All parameters are the same as in Fig.~\ref{fig:snapshot}.}
	\label{fig:rt_with_transient}
\end{figure}

\begin{figure}[] 
\centering
	\includegraphics[width = 0.9\linewidth]{{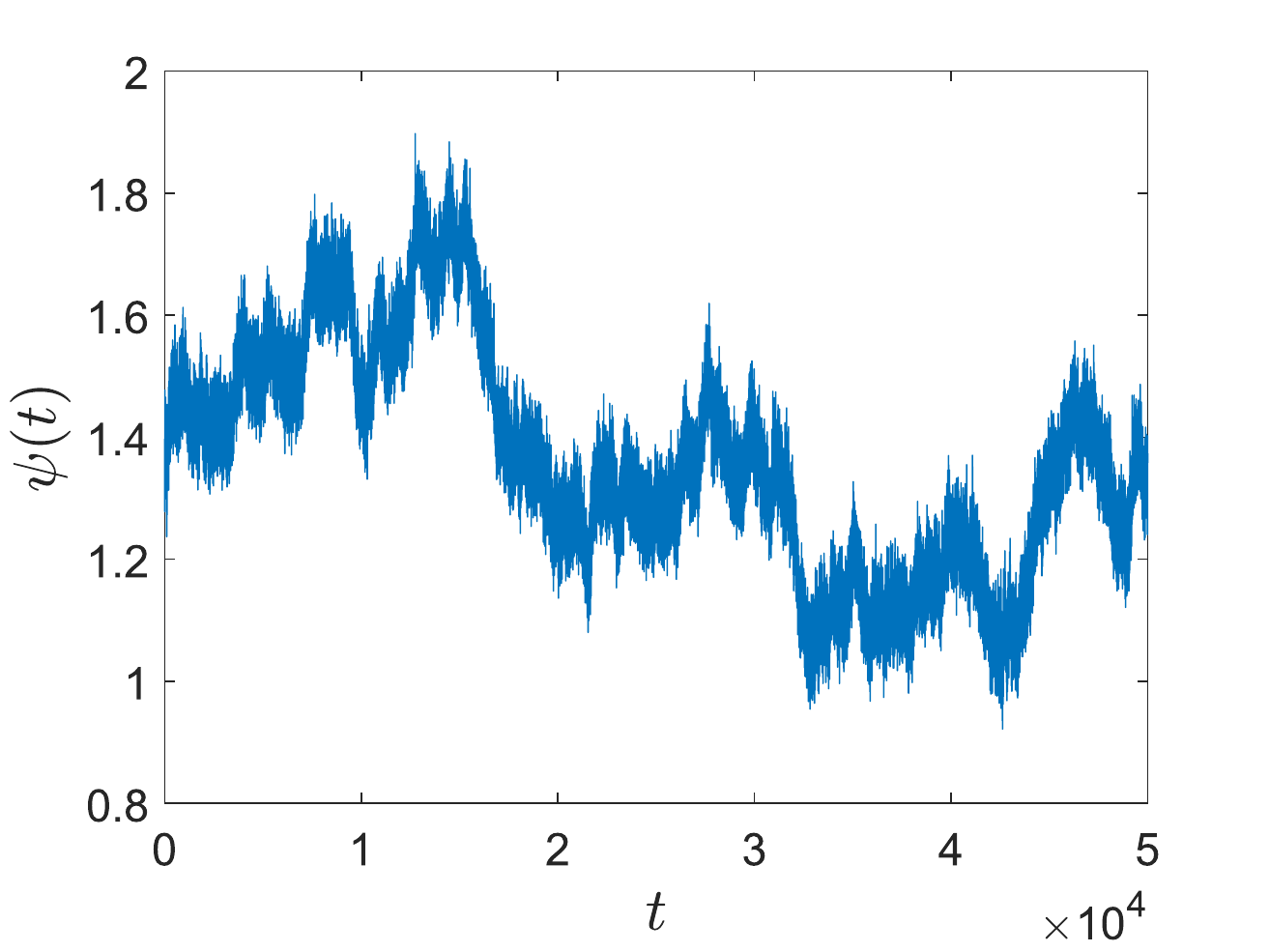}}
	\includegraphics[width = 0.9\linewidth]{{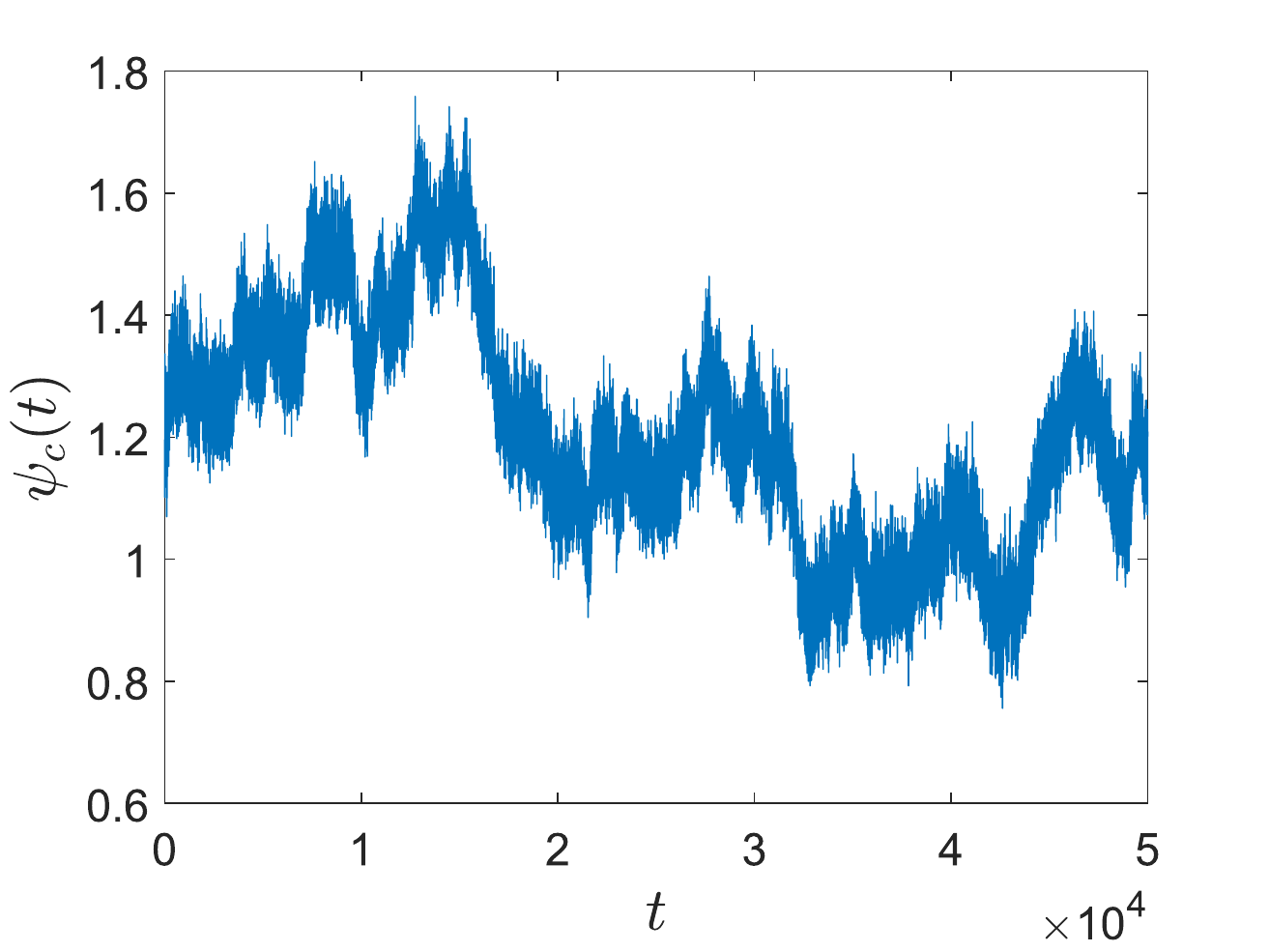}}
	\caption{Time series of the overall mean phase $\psi(t)$ defined in (\ref{eq:r0}) and of the mean phase $\psi_c$ of the synchronized cluster defined in (\ref{eq:rc}) obtained from a simulation of the Kuramoto-Sakaguchi model (\ref{eq:ks}) with $N = 160$ oscillators, exhibiting diffusive behaviour. All parameters are the same as in Fig.~\ref{fig:snapshot}.}
	\label{fig:psi}
\end{figure}

The effect of the non-collective behaviour of the rogue oscillators induces a seemingly stochastic behaviour of the order parameter $r$ shown in Figure~\ref{fig:rt_with_transient}. We show the temporal evolution of the order parameter from a random initial condition at $t=0$; after an initial transient the order parameter exhibits persistent fluctuations around a mean value with a constant variance of $4.16\times10^{-4}$ for $N=160$ and $4.42\times10^{-6}$ for $N=10,000$. The size of the fluctuations is $N$ dependent and we expect that for $N\to \infty$ the variance of the fluctuations approaches zero as demonstrated in Figure~\ref{fig:rt_stats} further down.  
Figure~\ref{fig:psi} shows that the overall mean phase $\psi(t)$ as defined in (\ref{eq:r0}) as well as the mean phase of the synchronized oscillators only, as defined further down in (\ref{eq:rc}), exhibit diffusive behaviour. We checked that their dynamics is (approximately) Brownian with a linear time dependency of the mean-square displacement (not shown). Finite-size induced diffusivity of the phase has been previously reported for a stochastic Kuramoto model with $\lambda=0$ and additive noise \cite{BertiniEtAl10, BertiniEtAl14,GiacominPoquet15,Lucon15,LuconPoquet15,LuconStannat16,Gottwald17}. Here the effect is entirely a deterministic finite-size effect as there is no driving noise in the Kuramoto-Sakaguchi model (\ref{eq:ks}). \\

\subsection{Dependency of fluctuations on the system size $N$}
\begin{figure}[]
\centering
\begin{subfigure}[]{\columnwidth}
\centering
	 \includegraphics[width = 0.9\linewidth]{{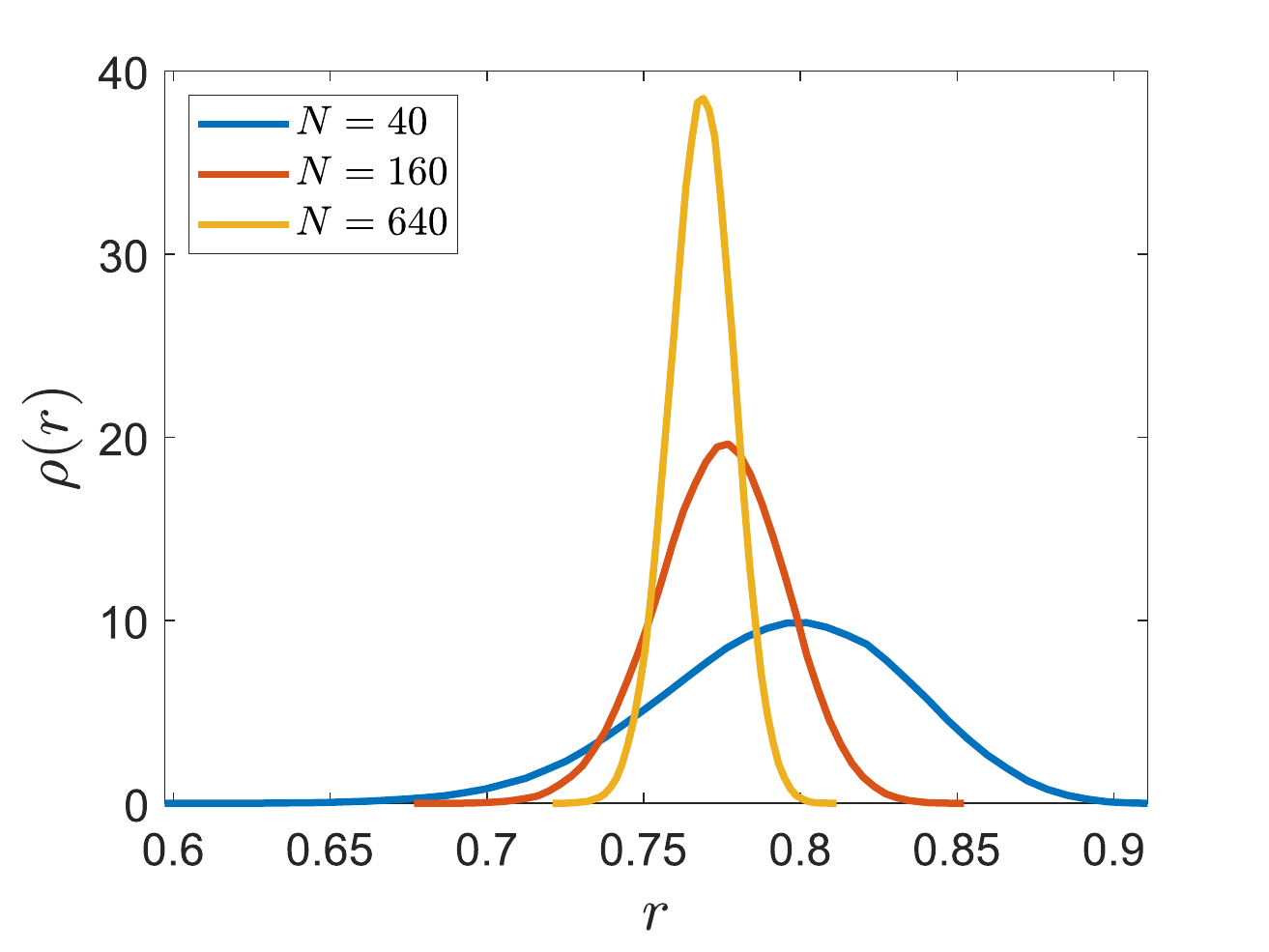}}
\caption{}
\end{subfigure}
\hfill
\begin{subfigure}[]{\columnwidth}
\centering
	\includegraphics[width = 0.9\linewidth]{{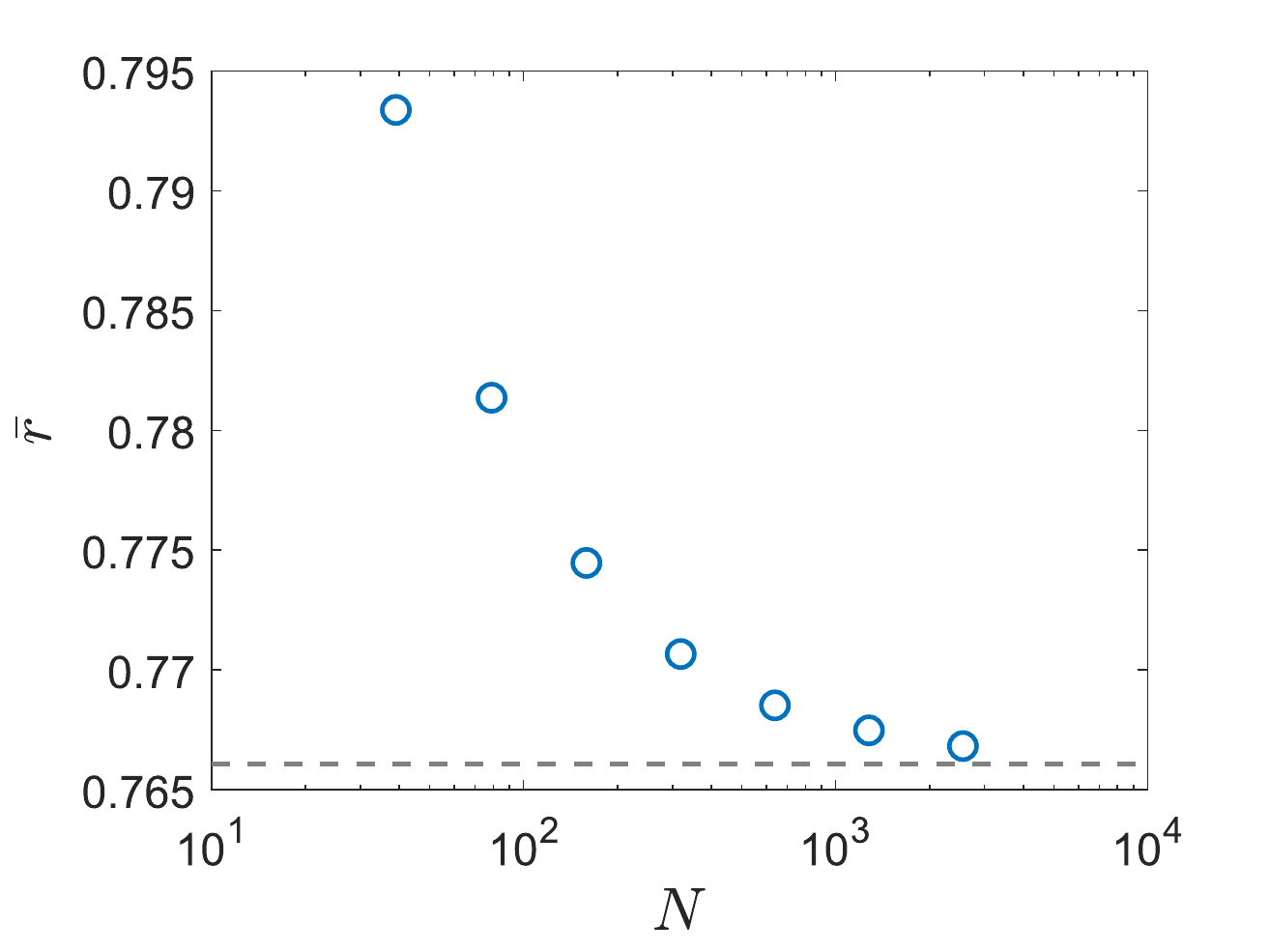}}
\caption{}
\end{subfigure}
\hfill
\begin{subfigure}[]{\columnwidth}
\centering
	\includegraphics[width = 0.9\linewidth]{{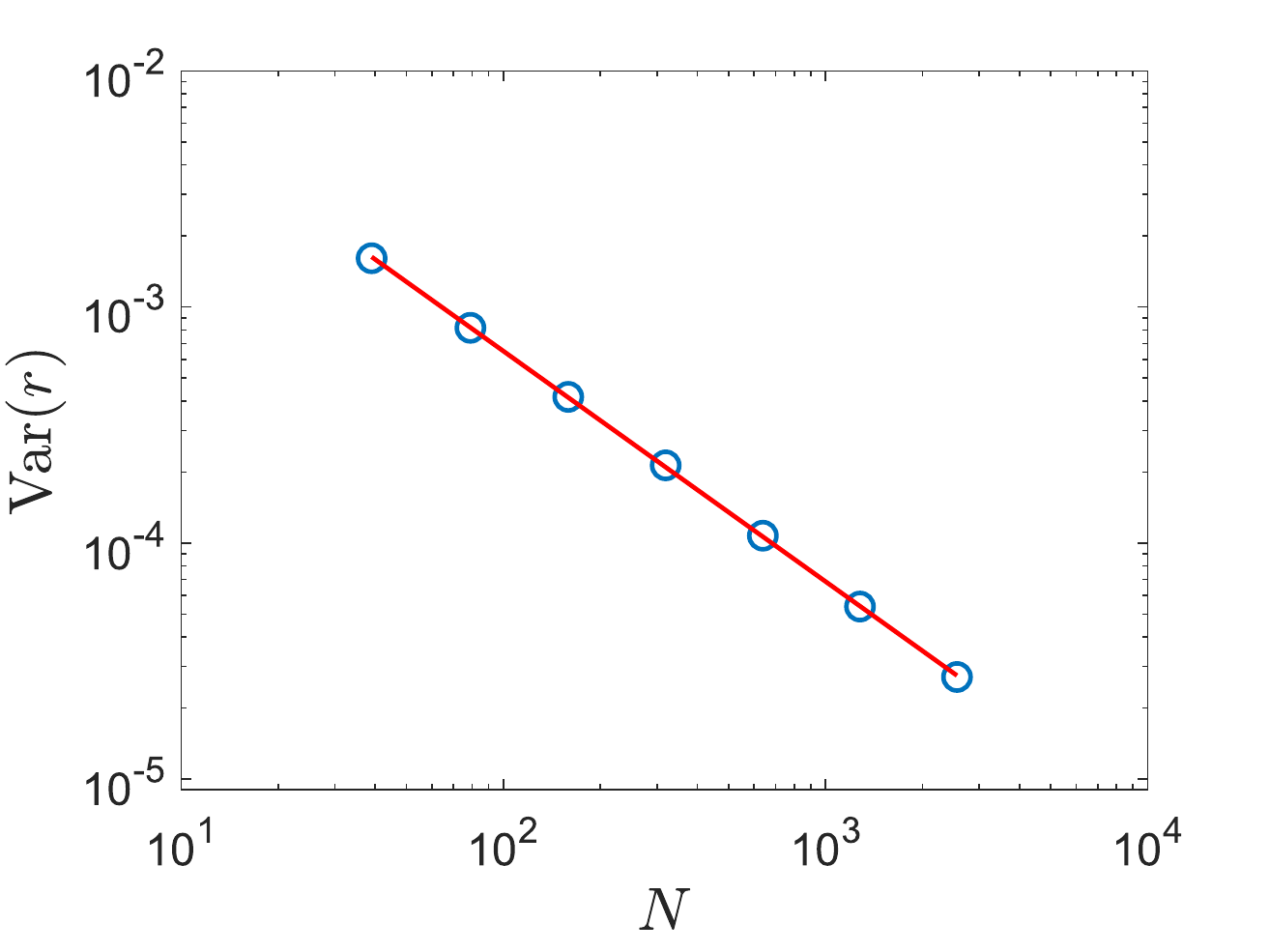}}
\caption{}
\end{subfigure}
\caption{(a): Empirical distribution of the order parameter $r(t)$ obtained from a single simulation of the Kuramoto-Sakaguchi model (\ref{eq:ks}) at different system sizes $N$. (b): Mean value of $r(t)$ for different system sizes $N$. The dashed line indicates the value at the corresponding thermodynamic limit, $\bar{r}_\infty$. (c): Variance of $r(t)$ for different system sizes $N$. The red line represents the best-fit suggesting a scaling law of $N^{-0.976}$. All other parameters are the same as in Fig.~\ref{fig:snapshot}.}
	\label{fig:rt_stats}
\end{figure}

To investigate the effect of finite system size, we simulate the system at increasing system size $N = \{40,80,160,320,640,1280,2560\}$ with other settings kept the same. Figure \ref{fig:rt_stats}a shows the distribution of $r(t)$ for varying system size. As $N$ increases, the distribution of $r(t)$ becomes, as expected, narrower with decreasing variance and its mean value approaches a fixed value $\bar r_\infty = 0.766$, as shown in Figures~\ref{fig:rt_stats}b-c. We find that $\bar r - r_{\infty} \sim N^{-0.859}$. The variance of $r(t)$ decreases towards 0 as $N$ increases with $\textrm{Var}[r] \sim N^{-0.976}$. The scaling of the variance with approximately with $1/N$ suggests an underlying Central Limit Theorem describing the finite-size fluctuations of the order parameter around the thermodynamic mean as a Gaussian process. \\

\begin{figure}[] 
\centering
	\includegraphics[width = 0.9\linewidth]{{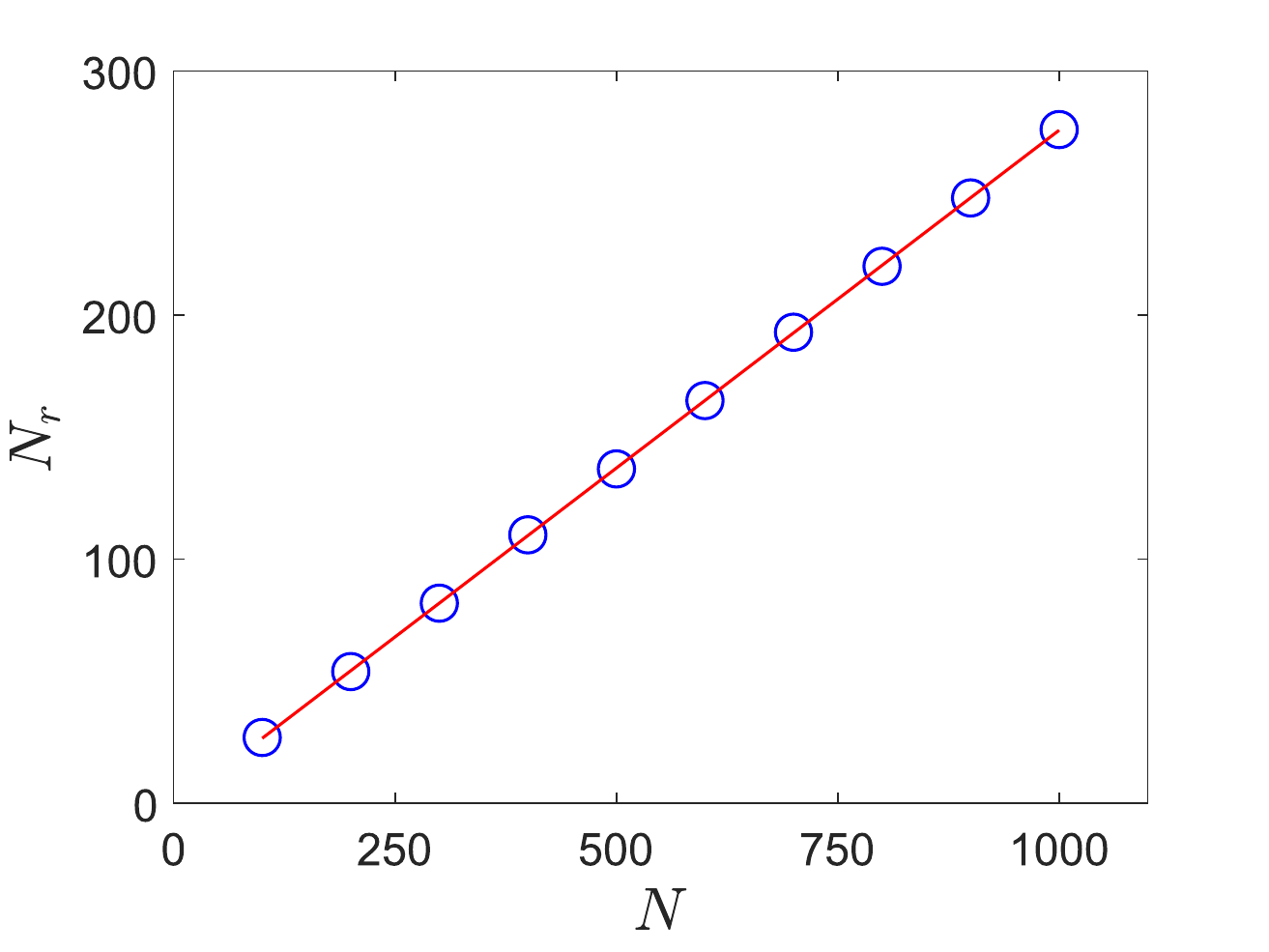}}
	\caption{The number of rogue oscillators $N_r$ as a function of the total number of oscillators $N$. The continuous line shows a best fit with slope $0.2767$ in good agreement with the thermodynamic limit relationship (\ref{eq:Nr}).  All other parameters are the same as in Fig.~\ref{fig:snapshot}.}
	\label{fig:Nr}
\end{figure}

Figure~\ref{fig:Nr} shows that the number of rogue oscillators $N_r$ very closely exhibits a linear dependency on the total number of oscillators $N$. The numerically estimated slope of $0.2767$ is very well approximated by the thermodynamic limit 
\begin{align}
\lim_{N\to \infty} \frac{N_r}{N} = \int_{|\omega-\Omega|>Kr} g(\omega) d\omega,
\label{eq:Nr}
\end{align}
where $r$ and $\Omega$ are the solutions of the mean-field consistency relations (\ref{eq:self_consistency_real}--(\ref{eq:self_consistency_imag}). 
For the parameters used in Figure~\ref{fig:Nr} we obtain $\lim_{N\to \infty} \frac{N_r}{N} = 0.2770$. 

\subsection{Dependency of fluctuations on the coupling strength $K$}
\begin{figure}[]
\begin{subfigure}[]{\columnwidth}
\centering
	\includegraphics[width = 0.9\linewidth]{{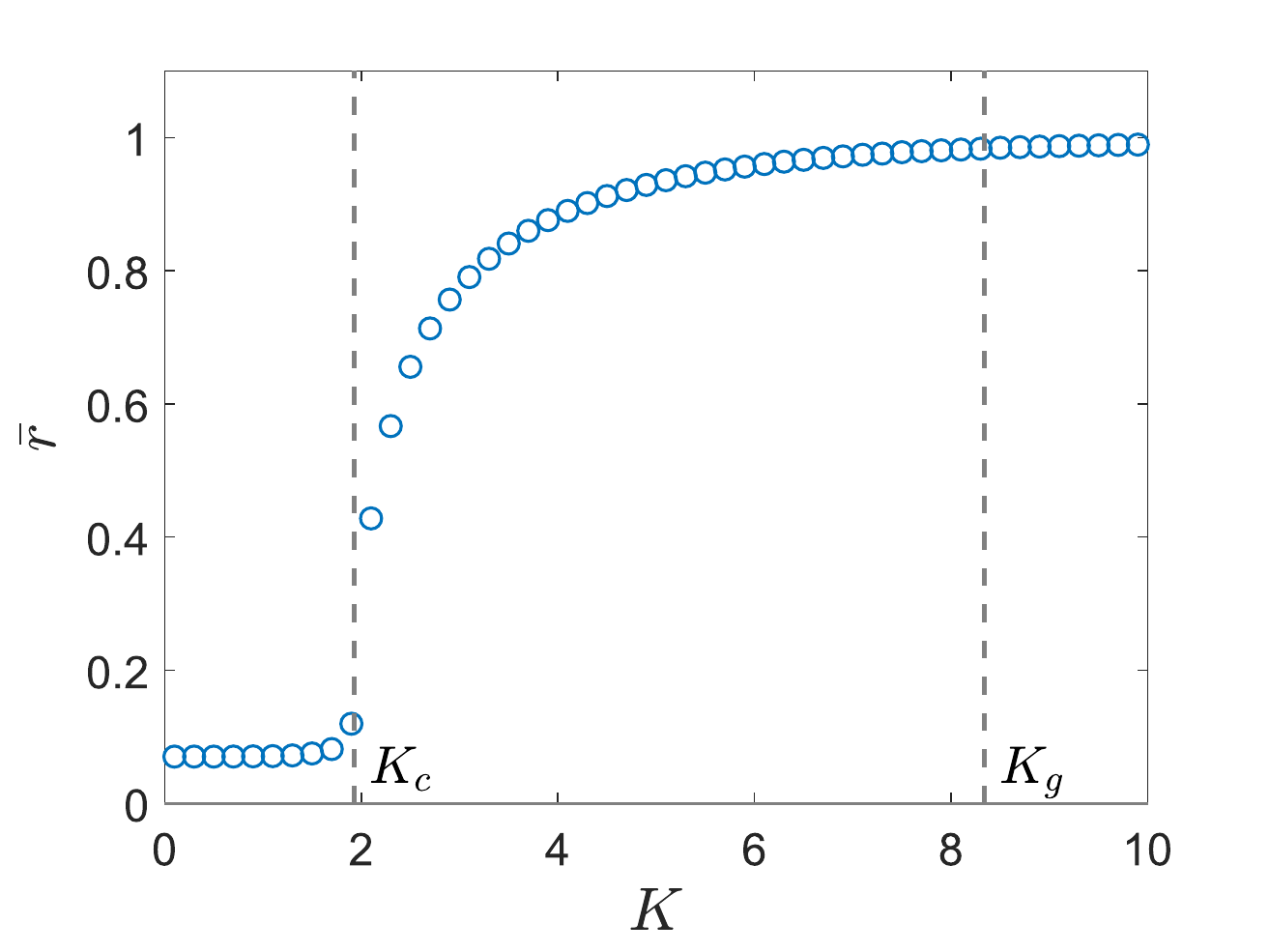}}\\
\caption{}
\end{subfigure}
\hfill
\begin{subfigure}[]{\columnwidth}
\centering
	\includegraphics[width = 0.9\linewidth]{{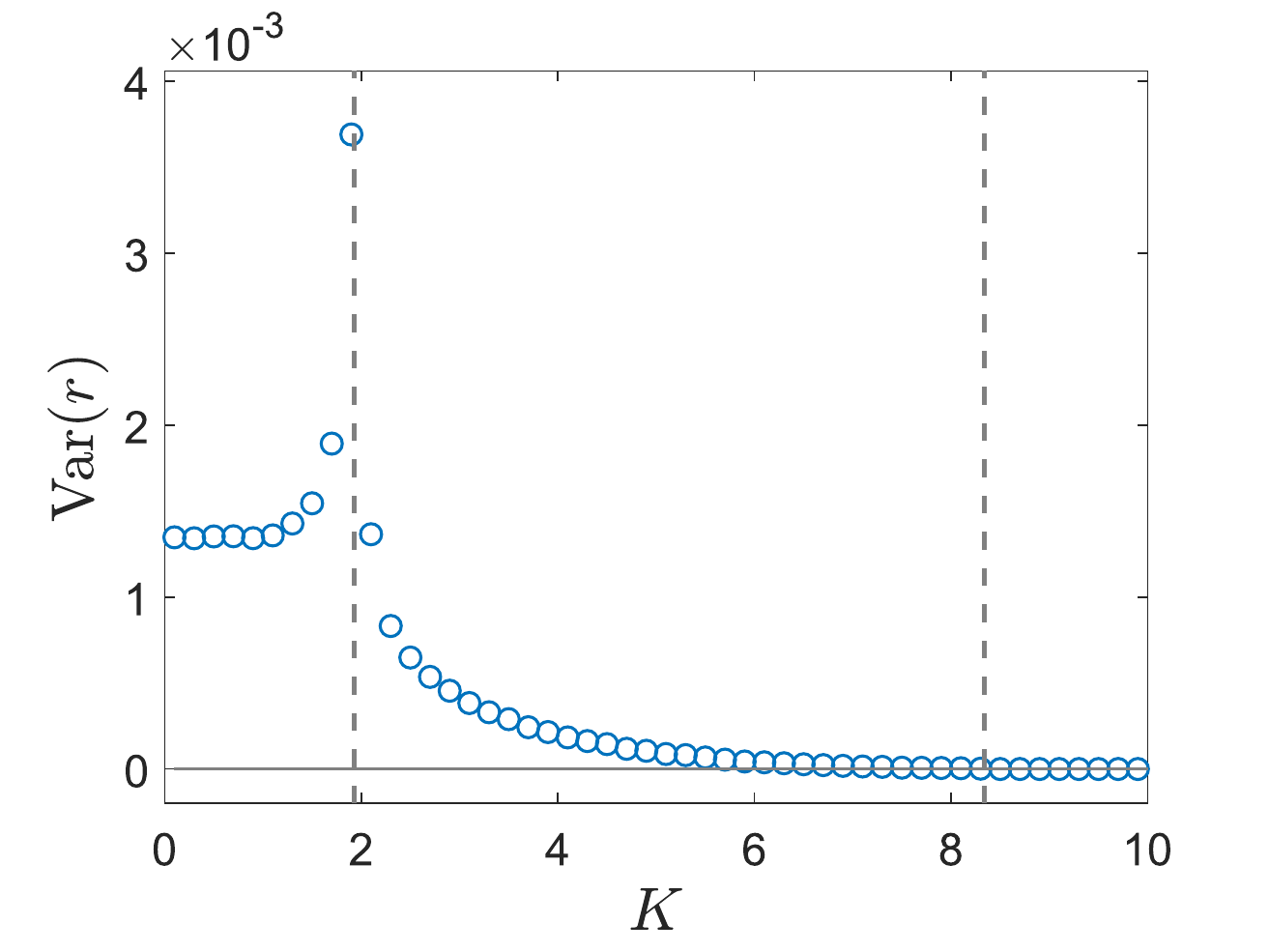}}\\
\caption{}
\end{subfigure}
\hfill
\begin{subfigure}[]{\columnwidth}
\centering
	\includegraphics[width = 0.9\linewidth]{{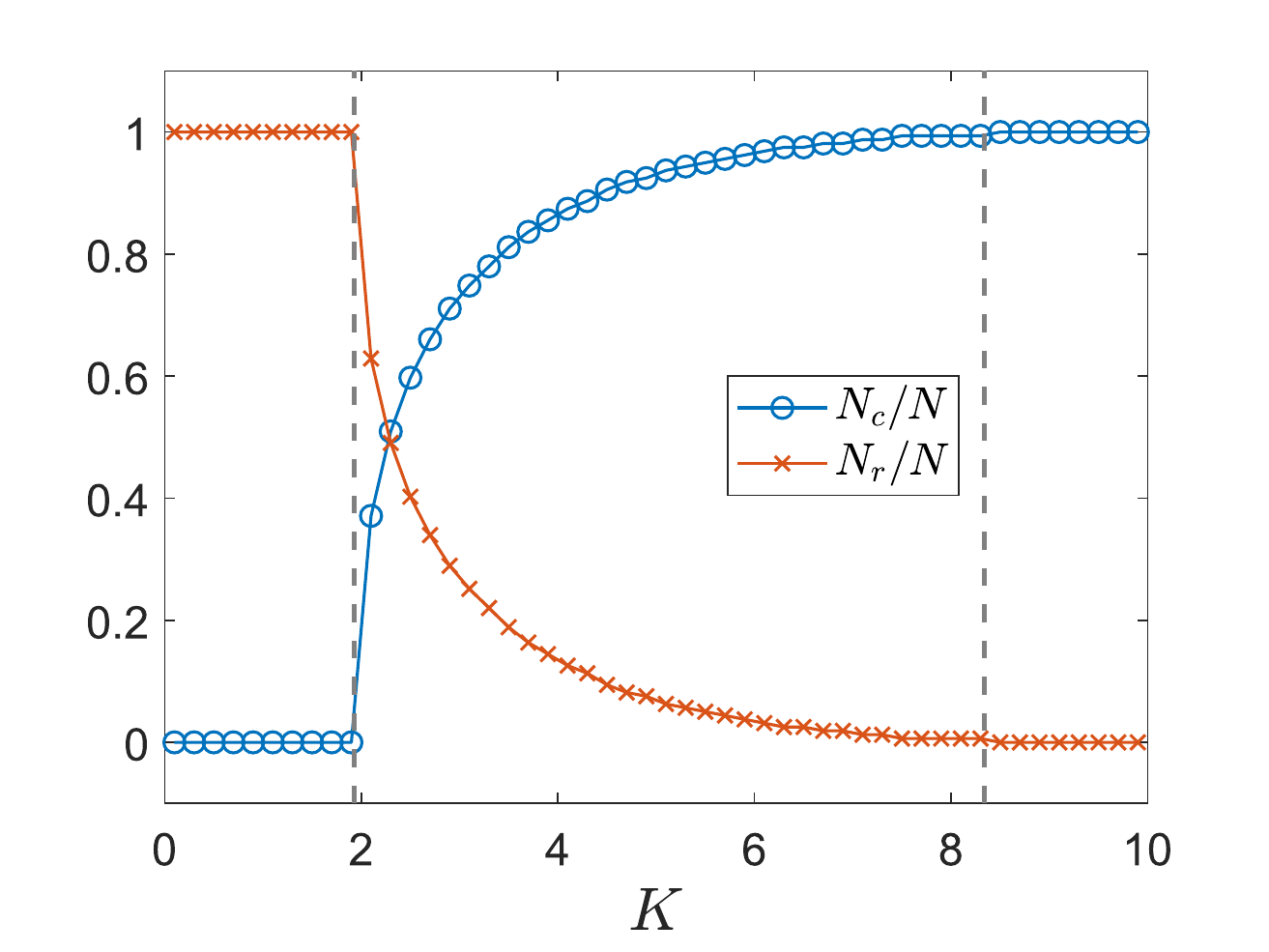}}
\caption{}
\end{subfigure}
	\caption{Transition from incoherence to synchronisation of the Kuramoto-Sakaguchi model (\ref{eq:ks}) with increasing coupling strength $K$ with $N=160$ for a Gaussian intrinsic frequency distribution $g(\omega)$ with mean zero and variance $1$. (a): Order parameter $\bar r$. (b): Variance of the order parameter, quantifying the size of fluctuations. (c): Number of entrained oscillators $N_c$, which are part of the collective synchronised cluster (circles, online blue) and number of non-entrained rogue oscillators $N_r$ (crosses, online red). Shown are relative numbers. The vertical lines mark the onset of partial synchronisation at $K_c=1.93$ and the onset of global synchronisation at $K_g=8.34$.}
	\label{fig:ks_diff_K}
\end{figure}

Figure~\ref{fig:ks_diff_K} shows the well-known transition from incoherence to synchronisation for increasing coupling strength $K$ for a fixed number of oscillators $N=160$. Figure~\ref{fig:ks_diff_K}a shows the order parameter $\bar r$ with the well known second order phase transition for Gaussian intrinsic frequencies from incoherence with $\bar r \sim 1/\sqrt{N}$ to partial synchronisation at $K_c = 1.93$, after which the synchronised cluster continues to grow in size until all oscillators become synchronised at $K_g = 8.34$. Figure~\ref{fig:ks_diff_K}b  shows the variance of the order parameter as a function of the coupling strength. The almost constant variance for small coupling strength around ${\rm{Var}}(r) = 1.35\times 10^{-3}$ corresponds to the almost random distribution of the $N$ uncoupled oscillators. The variance exhibits a singularity at $K=K_c$ and approaches zero for $K\ge K_g$. The singular behaviour of the variance is a well-studied phenomenon and known under the name of anomalous enhancement of fluctuations \cite{Daido87,Daido89,Daido90,HongEtAl07,Tang11,HongEtAl15}. Figure~\ref{fig:ks_diff_K}c shows the cardinalities of the synchronised cluster $\mathcal{C}$ and the set of rogue oscillators $\mathcal{R}$, labelled $N_c$ and $N_r$, respectively, with $N_c+N_r=N$. For $K<K_c$ all oscillators are rogue whereas for $K>K_g$ all oscillators are synchronised. The synchronised cluster steadily grows for increasing $K_c<K<K_g$.

\section{Stochastic model reduction}
\label{sec:sde}

The numerical results presented in Section~\ref{sec:num} suggest that the non-entrained rogue oscillators exert a stochastic forcing on the synchronised cluster. To develop a stochastic approximation of the Kuramoto-Sakaguchi model we hence aim to establish a closed evolution equation for the phases of the synchronised oscillators in which the driving force exerted by the rogue oscillators is parametrised by a stochastic process. The challenge is how to describe this stochastic process. We begin by separating the coupling terms which only involve the synchronised oscillators and those which contain the rogue oscillators, and write the Kuramoto-Sakaguchi model (\ref{eq:ks}) for $i\in \mathcal{C}$ as
\begin{align} 
	\dot{\theta_i}  = \omega_i-\Omega+ \frac{K}{N}\left(\sum_{j\in\mathcal{C}}\sin(\theta_j-\theta_i-\lambda)+\sum_{j\in\mathcal{R}}\sin(\theta_j-\theta_i-\lambda)\right).
	\label{eq:ks_S_C_sep}
\end{align}
The sum over the rogue oscillators can be written as
\begin{align*} 
	&\frac{1}{N}\sum_{j\in\mathcal{R}}\sin(\theta_j-\theta_i-\lambda)\\
	&\qquad =\frac{1}{N}\sum_{j\in\mathcal{R}}\sin(\theta_j-\psi_c+\lambda+\psi_c-\theta_i-2\lambda)\\
	&\qquad =\cos(\psi_c-\theta_i-2\lambda)\, S(t) + \sin(\psi_c-\theta_i-2\lambda) \, C(t),
\end{align*}
where we introduced the mean phase of the synchronised cluster $\psi_c$ and where we define
\begin{align} 
	S(t)&=\frac{1}{N}\sum_{j\in\mathcal{R}}\sin(\theta_j-\psi_c+\lambda)= \frac{N_r}{N} r_r\sin(\psi_r-\psi_c+\lambda)
	\label{eq:S}
	\\
	C(t)&=\frac{1}{N}\sum_{j\in\mathcal{R}}\cos(\theta_j-\psi_c+\lambda)= \frac{N_r}{N} r_r\cos(\psi_r-\psi_c+\lambda),
	\label{eq:C}
\end{align}
where we defined the mean-field variables pertaining to the synchronized oscillators in $\mathcal{C}$ 
\begin{align}
r_ce^{i\psi_c} = \frac{1}{N_c}\sum_{j\in\mathcal{C}}e^{i\theta_j}
\label{eq:rc}
\end{align}
and those pertaining to the rogue oscillators in $\mathcal{R}$
\begin{align}
r_re^{i\psi_r} = \frac{1}{N_r}\sum_{j\in\mathcal{R}}e^{i\theta_j}.
\label{eq:rr}
\end{align}
We remark that the particular choice of writing the trigonometric functions in (\ref{eq:S})-(\ref{eq:C}) was motivated by the form of the invariant density of the rogue oscillators (\ref{eq:rho_rogue}) and allows for a convenient averaging (cf. (\ref{eq:Sbar})-(\ref{eq:Cbar})). The overall influence of the rogue oscillators on each synchronised oscillator $\theta_i$ is now captured by $S(t)$ and $C(t)$, while the remaining terms are expressed entirely in terms of the synchronised oscillators, and we arrive at
\begin{align} 
	\dot{\theta_i}  &= \omega_i-\Omega+ \frac{K}{N}\sum_{j\in\mathcal{C}}\sin(\theta_j-\theta_i-\lambda)
	\nonumber
	\\
	&+K\cos \tilde\theta_i \; S(t) + K\sin \tilde\theta_i  \; C(t),
\label{eq:ks_SC}
\end{align}
where we defined for compactness 
\begin{align*}
\tilde\theta_i = \psi_c-\theta_i-2\lambda. 
\end{align*}
We first establish the average effect of the rogue oscillators and calculate the means $\langle S \rangle$ and $\langle C \rangle$ of the rogue oscillator drivers $S(t)$ and $C(t)$. We follow here the averaging procedure proposed in our previous work \cite{YueEtAl20}. 

In the co-rotating frame of the synchronized cluster, the synchronized oscillators are stationary whereas the rogue oscillators rotate (cf. Figure~\ref{fig:snapshot}). Envoking ergodicity we can equate the temporal averages $\langle S \rangle$ and $\langle C \rangle$ by averages over the stationary density function of the rogue oscillators (\ref{eq:rho_rogue}). Averaging equation (\ref{eq:S}) for $S(t)$ over the invariant densities of the rogue oscillators (\ref{eq:rho_rogue}) becomes
\begin{align} 
\langle S\rangle_t &\approx  \frac{1}{N} \sum_{j \in \mathcal{R}}\, \int_{0}^{2\pi}\sin(\theta_j-\psi_c-\lambda)\rho(\theta_j;\omega)d\theta_j \nonumber \\
&=\frac{1}{N}\cos(\psi-\psi_c)\sum_{j\in \mathcal{R}}k_j,
\label{eq:Sbar}
\end{align}
where
\begin{align} 
\label{eq:k_j}
k_j=\frac{\omega_j-\Omega}{Kr}\left(1-\sqrt{1-\frac{K^2r^2}{(\omega_j-\Omega)^2}}\right),
\end{align}
and $\psi$ denotes the overall phase of all oscillators in the frame of reference rotating with the mean frequency $\Omega$. Similarly, averaging (\ref{eq:C}) yields
\begin{align} 
\langle C\rangle_t =-\frac{1}{N}\sin(\psi-\psi_c)\sum_{j\in \mathcal{R}}k_j \approx 0,
\label{eq:Cbar}
\end{align}
since the overall phase $\psi$ (see \eqref{eq:r0}) and the phase of the synchronized cluster $\psi_c$ are close for sufficiently large coupling strength. We remark that for $\lambda=0$ we have  $\langle S\rangle_t =\langle C\rangle_t=0 $. In practice we estimate $\langle S\rangle_t $ and $\langle C\rangle_t $ numerically from a long time trajectory, which yields $\langle S\rangle_t=0.148$ and $\langle C\rangle_t=-0.0236$, see Section~\ref{sec:num} for the parameters used in the numerical simulations. These numerically estimated values are well approximated by the analytical expressions (\ref{eq:Sbar}) and (\ref{eq:Cbar}) which yield $\langle S\rangle_t=0.144$ and $\langle C\rangle_t=-0.0233$. To evaluate the analytical expressions (\ref{eq:Sbar}) and (\ref{eq:Cbar}) the constant values of $r$ and $\Omega$ for the fixed finite system size $N$ are required. These would only be available from numerical simulations. To remain purely analytical we use instead the values corresponding to the thermodynamic limit with $r=0.766$ and $\Omega=-1.706$, which are calculated via the self-consistency relations (\ref{eq:self_consistency_real}) and (\ref{eq:self_consistency_imag}) of the mean-field theory. We remark that instead of using the finite size expressions (\ref{eq:Sbar}) and (\ref{eq:Cbar}) we could consider their thermodynamic limit by averaging over the frequency distribution $g(\omega)$. This yields $\langle S\rangle_t=0.149$ and $\langle C\rangle_t=-0.0240$ which is also close to the numerically observed values.\\

\begin{figure}[] 
	\includegraphics[width=0.9\linewidth]{{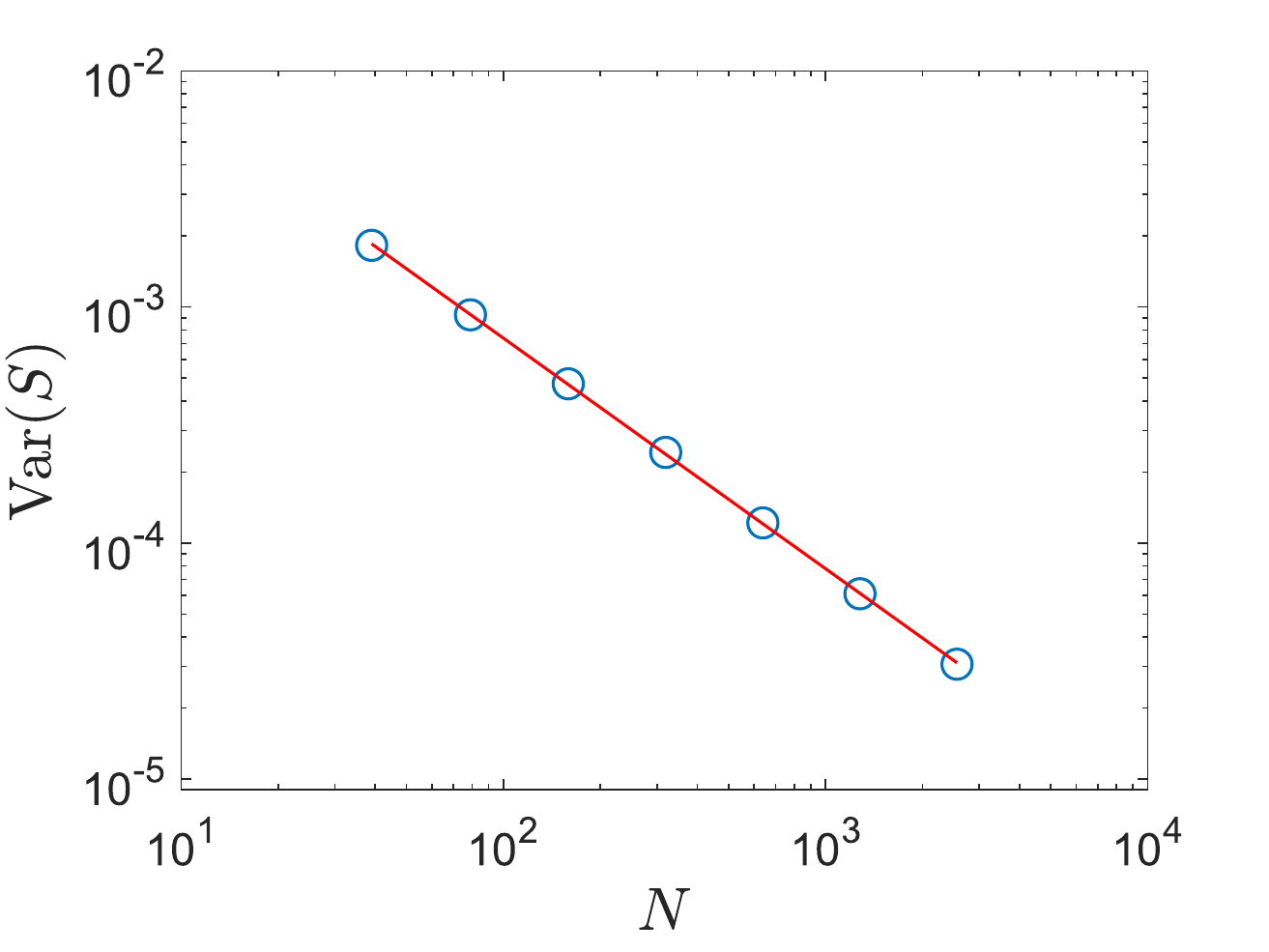}}
	\includegraphics[width=0.9\linewidth]{{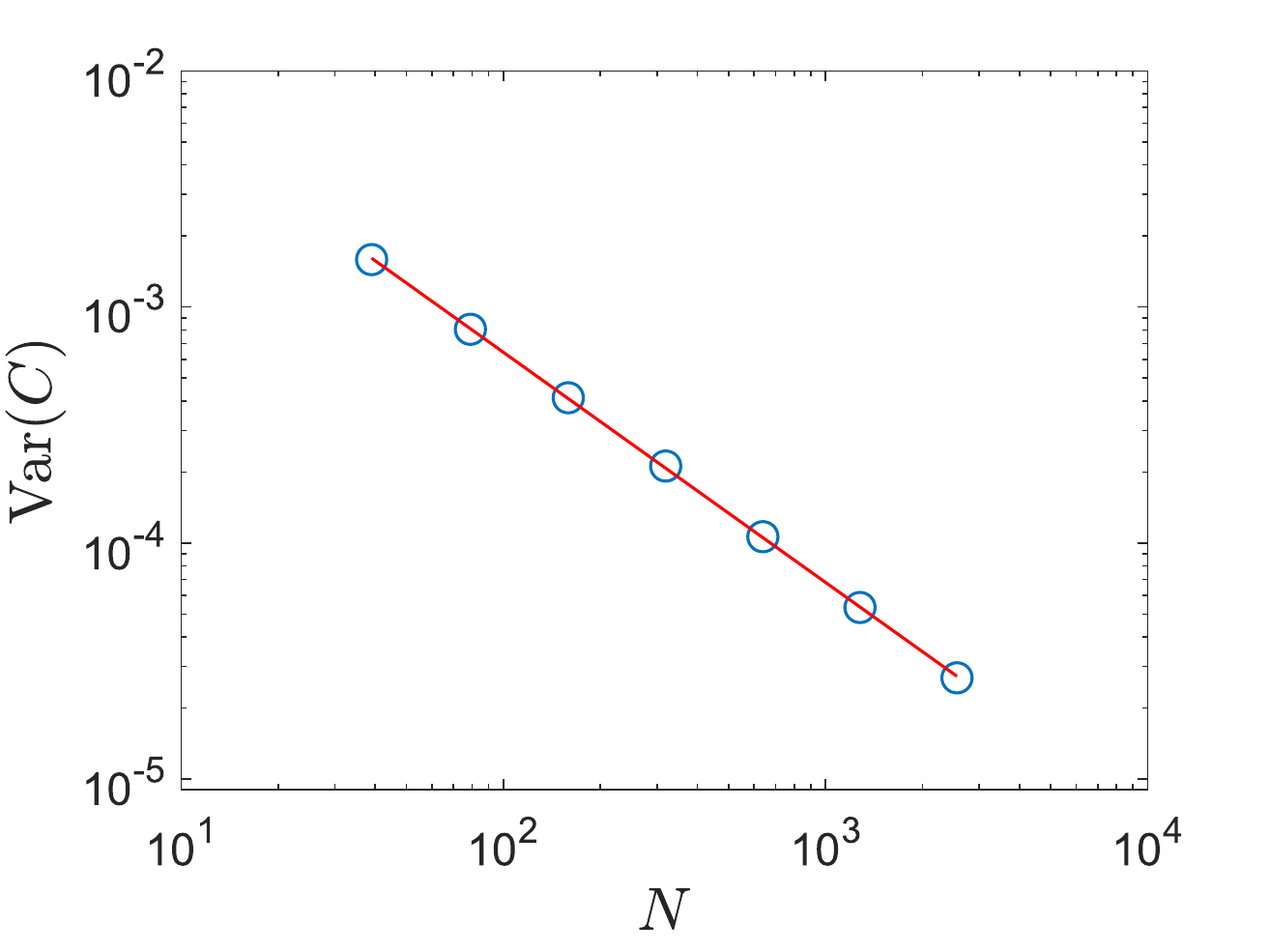}}
	\caption{Scaling of the variance of $S$ and $C$ with varying system size $N$ as calculated from a single trajectory of the Kuramoto-Sakaguchi model (\ref{eq:ks}). The continuous line is a linear best fit indicating a scaling with $N^{-0.977}$ and $N^{-0.975}$ for $S$ and $C$, respectively.}
	\label{fig:var_SC}
\end{figure}

Figure~\ref{fig:rho_SC} shows the empirical histogram of $S$ and $C$ obtained from a single long simulation of the Kuramoto-Sakaguchi model (\ref{eq:ks}) for the same parameter setting as in Section~\ref{sec:num} with $N=160,$ $K=3$, $\lambda=\pi/4$ and $ g(\omega)\sim\mathcal{N}(0,1)$. Whereas the mean values are well approximated by the averaging procedure \cite{YueEtAl20} described above, $S(t)$ and $C(t)$ experience significant fluctuations. It is clearly seen that the distribution of these fluctuations is Gaussian. In Figure~\ref{fig:var_SC} we show that the variance of $S$ and of $C$ scales approximately as $1/N$. This suggests that the scaled mean-subtracted variables
\begin{align}
\xi_t &= \sqrt{N}\left( S(t)  - \langle  S \rangle_t \right) + o(\frac{1}{\sqrt{N}})
\nonumber
\\
\zeta_t &= \sqrt{N}\left( C(t) - \langle  C \rangle_t \right)+ o(\frac{1}{\sqrt{N}})
\label{eq:xi}
\end{align}
are Gaussian processes that are defined entirely in terms of their mean and covariance functions 
\begin{align}
R_{ab}(\tau) = {\rm{cov}}(a(t),b(t+\tau)),
\label{eq:R}
\end{align}
for $a$ and $b$ being either $\xi$ or $\zeta$. 
Indeed, Figure~\ref{fig:covKM} shows that correlations decay in time with increasing system size $N$, and that the covariance functions of the scaled variables $\xi_t$ and $\zeta_t$ converge in the limit of $N\to \infty$. Note that the covariance functions of $\xi_t$ and $\zeta_t$ do not scale with $N$. 

The oscillations of the covariance functions for fixed finite $N$ are persistent features which do not decay for increasing length of the time series. To allow for a stochastic description of the fluctuations $\xi$ and $\zeta$ we require integrable covariance functions which decay in time. The dominance of the non-decaying large time correlations for small values of $N$ illustrates that the effective stochastic dynamics implied by \eqref{eq:xi} and the implied central limit theorem requires a sufficiently large number of oscillators $N$.   
We remark that technically one should consider the scaling with respect to the number of rogue oscillator $N_r$ rather than with the total number of oscillators $N$. However, since $N_r$ scales linearly with $N$ (cf Figure~\ref{fig:Nr} and equation (\ref{eq:Nr})) this notational simplification is justified.


\subsection{Approximation of $S(t)$ and $C(t)$ by a $2$-dimensional Ornstein-Uhlenbeck process}
\begin{figure}[]
\centering
	\includegraphics[width = 0.45\linewidth]{{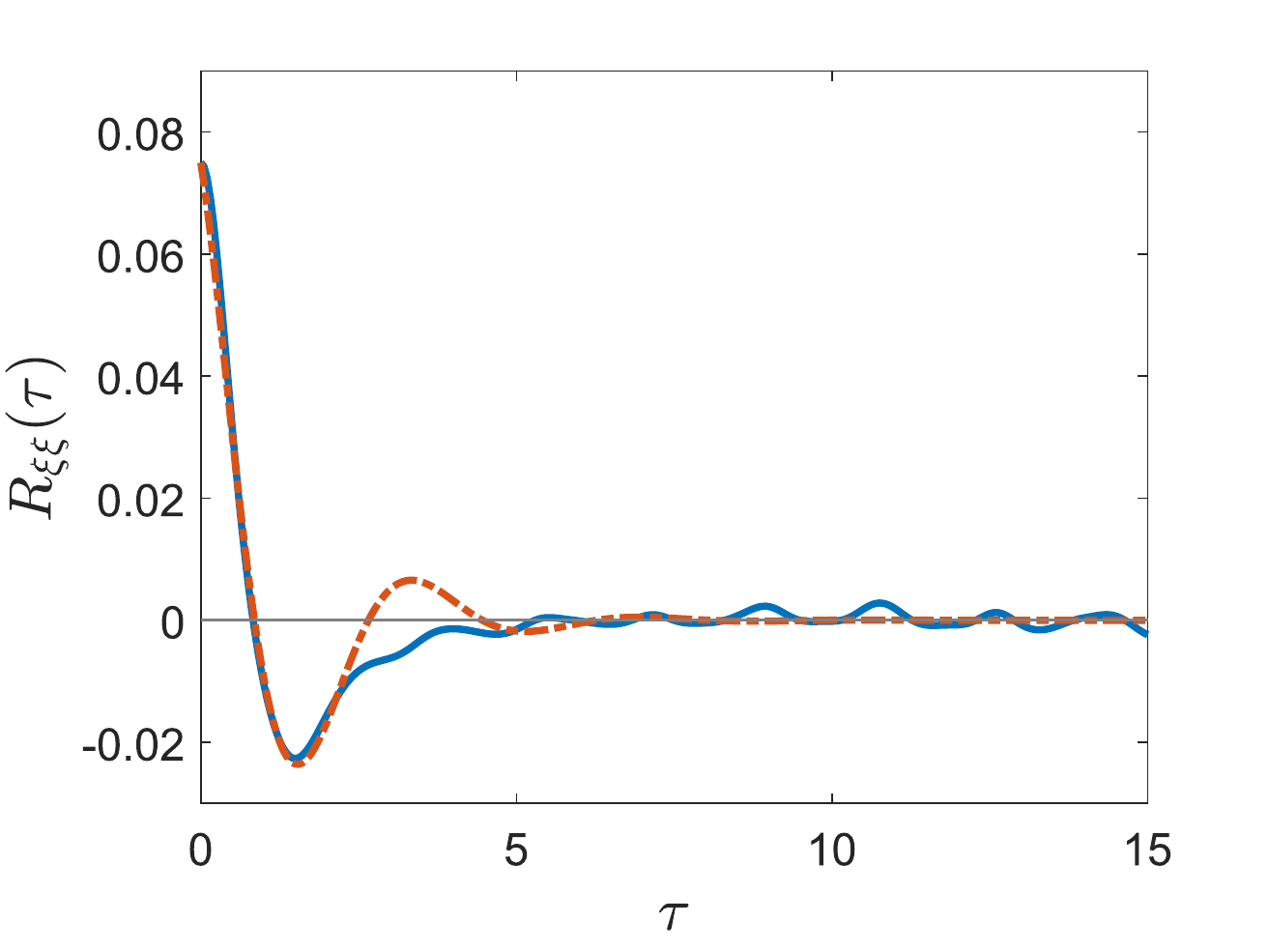}}
	\includegraphics[width = 0.45\linewidth]{{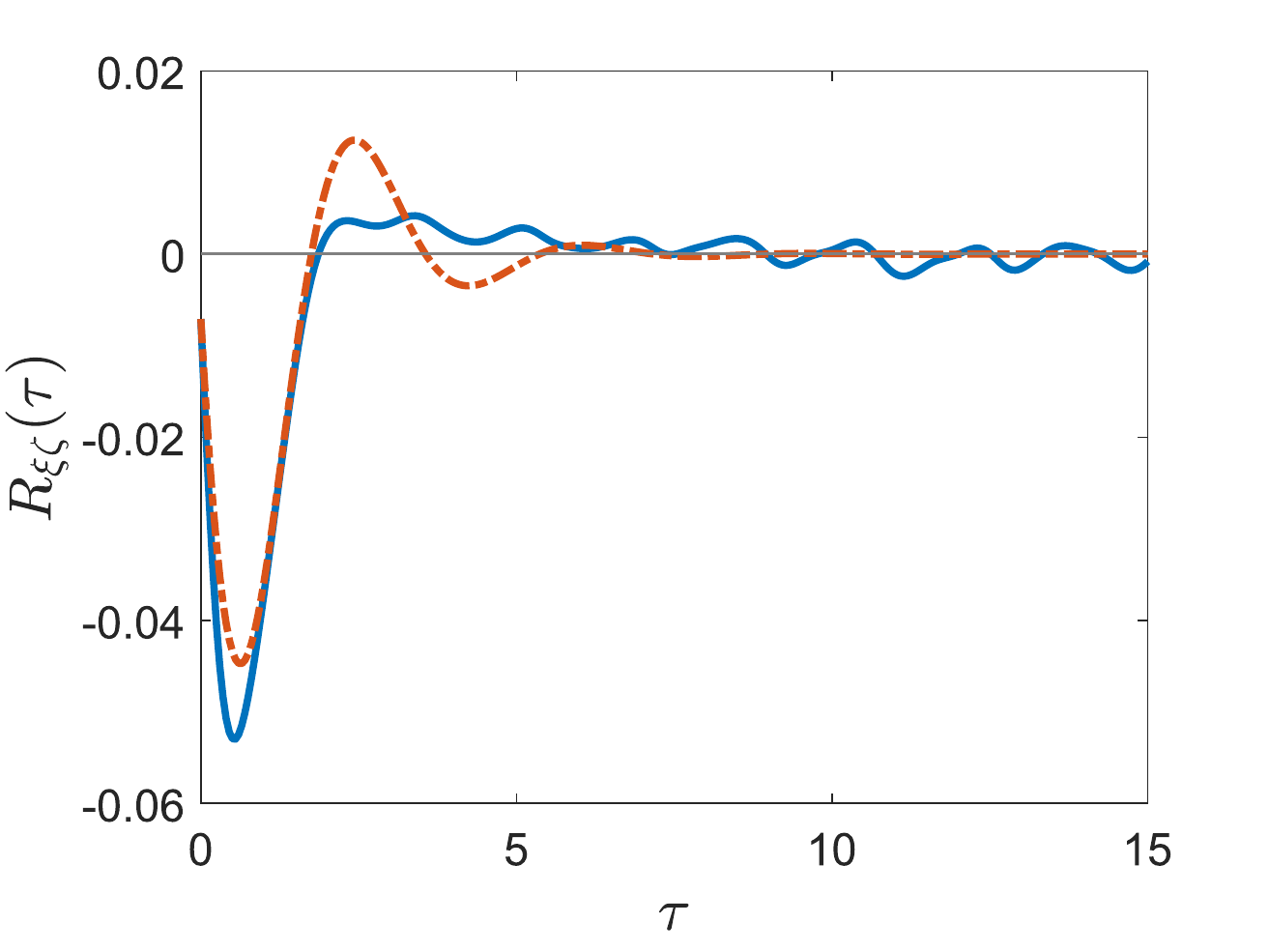}}\\
	\includegraphics[width = 0.45\linewidth]{{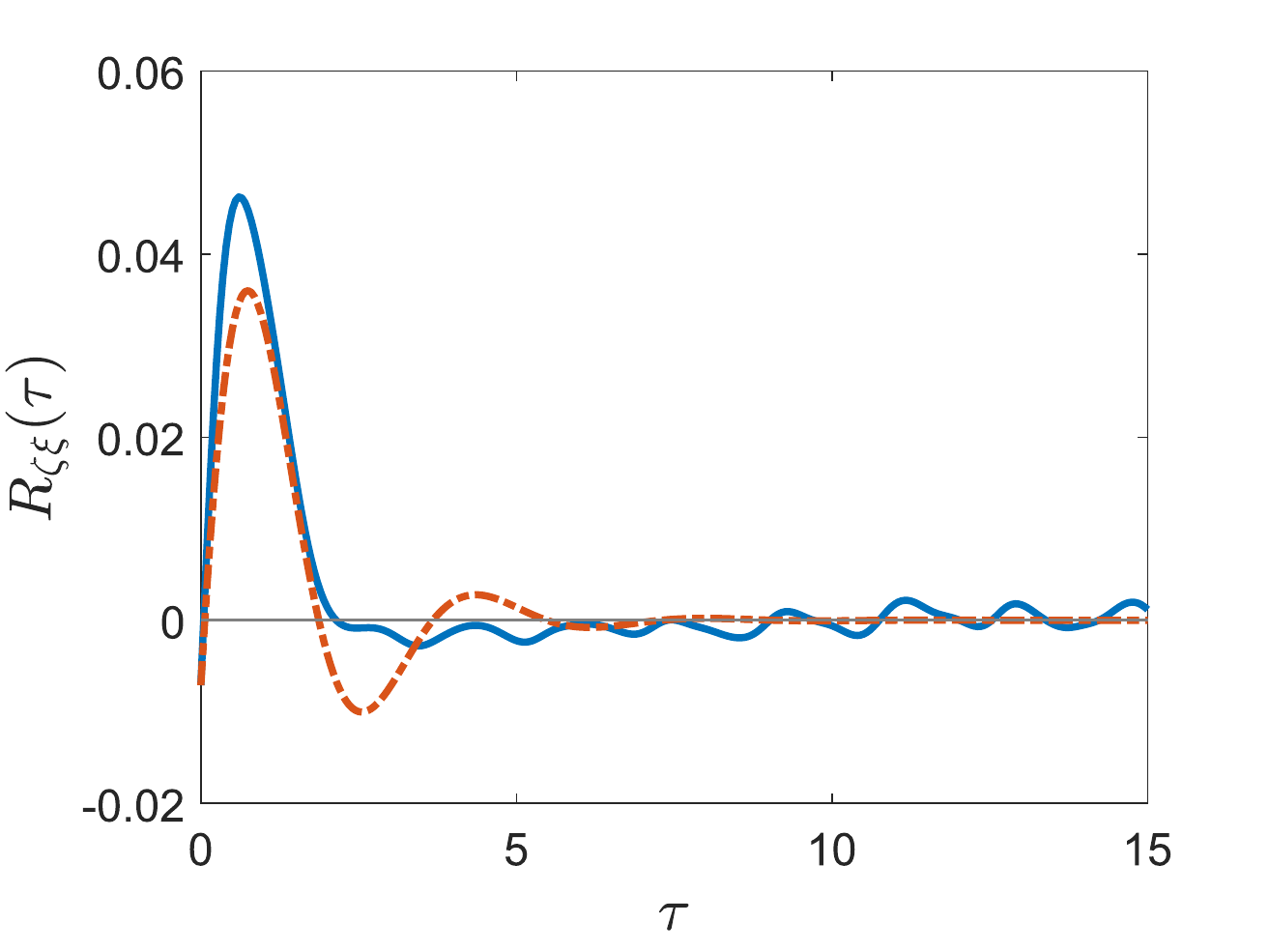}}
	\includegraphics[width = 0.45\linewidth]{{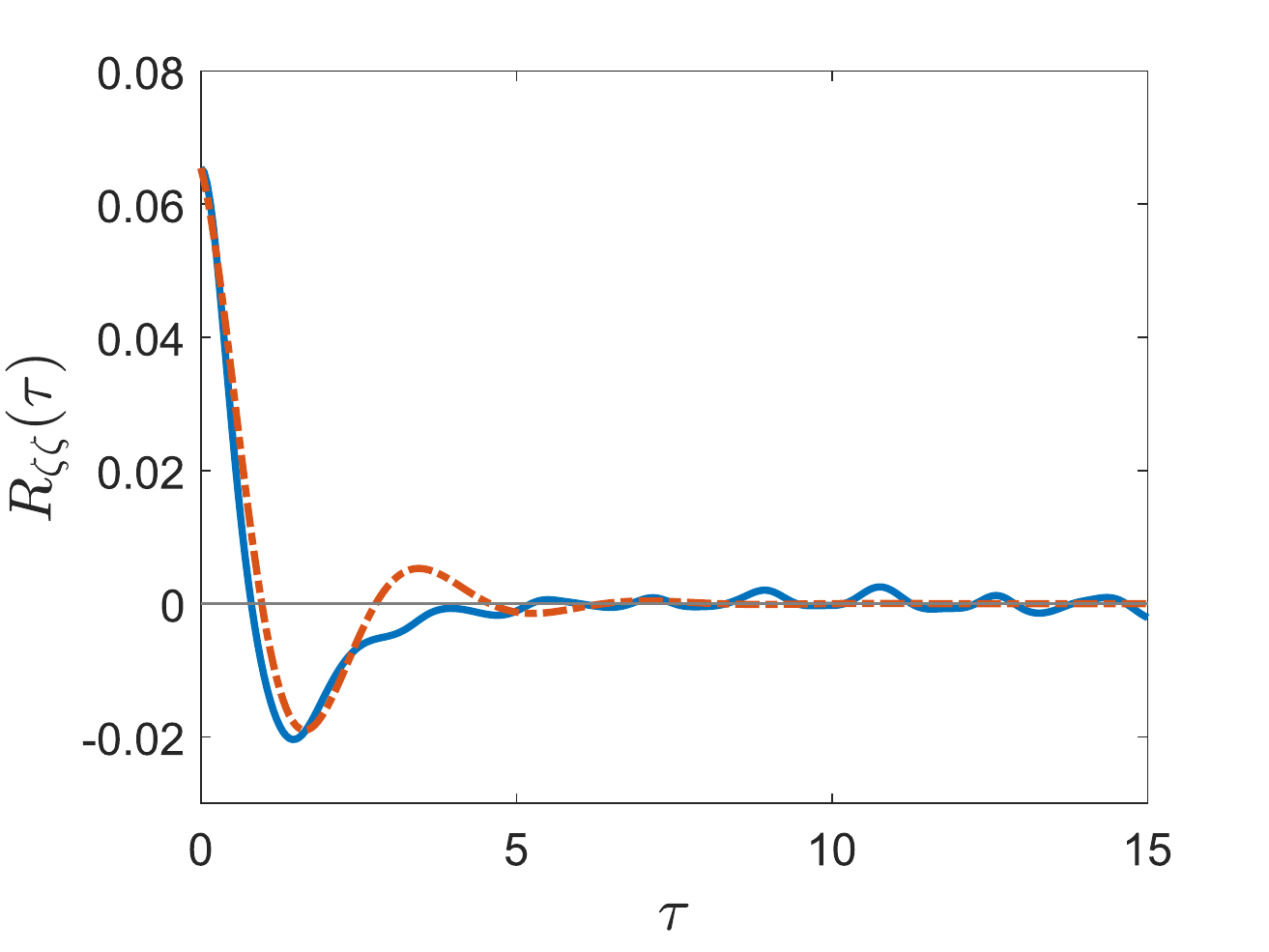}}
	\caption{Covariance functions of $\xi(t)$ and $\zeta(t)$ of the Kuramoto-Sakaguchi model (\ref{eq:ks}) with $N=160$ (continuous lines, online blue), together with the covariance function of a fitted two-dimensional Ornstein-Uhlenbeck process (dashed lines, online red).} 
	\label{fig:covOU}
\end{figure}

To determine an explicit stochastic process that can be used to generate realisations of this Gaussian process, we approximate the Gaussian process $Z_{\rm{KS}}=(\xi_t,\zeta_t)^T$ by an easily computable two-dimensional Ornstein-Uhlenbeck process
\begin{align}
dZ = \left( -\Gamma Z + \Upsilon Z\right)  dt + \Sigma \, dB_t
\label{eq:OU}
\end{align}
with two-dimensional Brownian motion $B_t=(B_1,B_2)^T$. It is well-known that stationary Gauss-Markov processes can be expressed as solutions of Ornstein-Uhlenbeck processes \cite{Ito84,GoldysEtAl16}. We consider for simplicity a diagonal matrix $\Gamma=\gamma\, \I$ and skew-symmetric rotation matrix $\Upsilon$ with entries $\upsilon_{12} = -\upsilon_{21}= \upsilon$ and $\upsilon_{11}=\upsilon_{22}=0$ and a symmetric diffusion matrix $\Sigma$ with entries $\sigma_{ij}$ for $i,j=1,2$. The associated covariance function is given by
\begin{align}
R^{(\rm{OU})}(\tau) = \langle e^{-(\Gamma - \Upsilon)\tau}Z_0 Z_0^T \rangle_{\rm{OU}},
\label{eq:ROU}
\end{align}
where the $Z_0$ are random variables drawn from the stationary density of the OU process and the angular brackets $\langle \cdot\rangle_{\rm{OU}}$ denote the average over that density. We provide explicit expressions of the covariance function $R^{(\rm{OU})}(\tau)$ in the Appendix. The parameters of the Ornstein-Uhlenbeck process \eqref{eq:OU} are determined such that its mean is zero and its covariance function \eqref{eq:ROU} best matches the observed covariance function (\ref{eq:R}) of $Z_{\rm{KS}}=(\xi_t,\zeta_t)$ associated with the full deterministic Kuramoto-Sakaguchi model. We perform a nonlinear least-square optimisation minimising the objective function
\begin{align}
E(\gamma, \upsilon,\Sigma) 
= \sum_{a,b} \,\int_{t_{\rm{min}}}^{t_{\rm{max}}} &
||R_{ab}(t)-{R_{ab}^{(\rm{OU})}}(t)||^2 dt \nonumber \\
&
 + \beta \left( ||R_{ab}(0)-{R_{ab}^{(\rm{OU})}}(0)||^2 \right) ,
\label{eq:LSQ}
\end{align}
where the sum goes over all entries of the covariance matrix. We choose $\beta=1,000$ to enforce that the variances of $S$ and $C$ are reproduced. For the parameter values used in Section~\ref{sec:num} with $K=3$, $\lambda=\pi/4$, $g(\omega) \sim \mathcal{N}(0,1)$ and $N=160$, the fitting yields  $\gamma=0.727\pm 0.0331, \upsilon=1.739\pm 0.0331$ and $\sigma_{11}=0.374\pm0.0069, \sigma_{12}=\sigma_{21}=0.0090\pm 0.0011, \sigma_{22}=0.271\pm 0.0084$ with $95\%$ confidence intervals. For different parameter values of the Kuramoto-Sakaguchi model (\ref{eq:ks}), the parameter values of the OU process will be different. Figure~\ref{fig:covOU} shows the four entries of the covariance matrix of $Z_{\rm{KS}}=(\xi_t,\zeta_t)$ together with the best fit of the approximating OU process. The fit is reasonable for $\tau\lesssim 2$.

\begin{figure}[] 
\centering
	\includegraphics[width=0.9\linewidth]{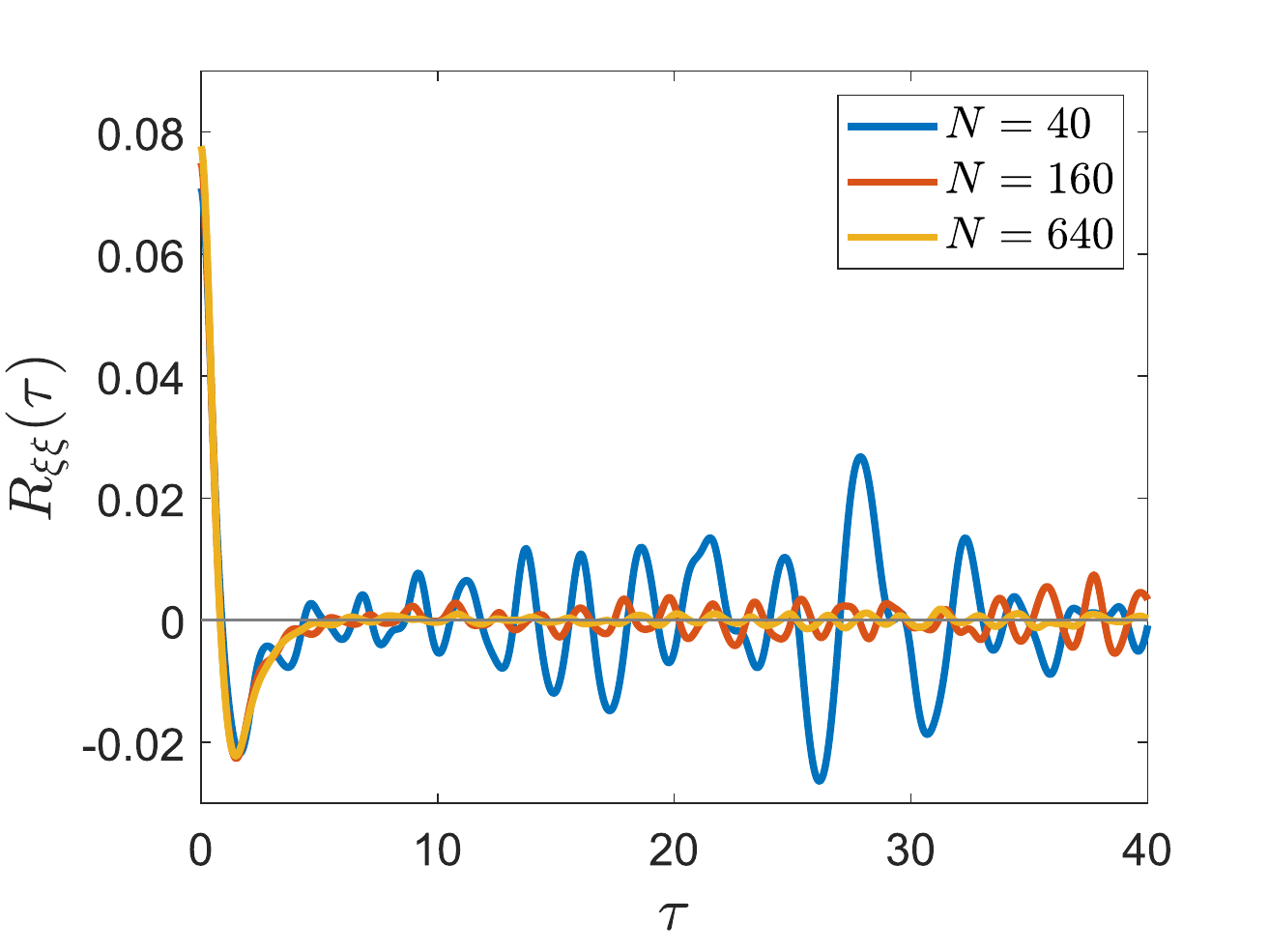}
	\includegraphics[width=0.9\linewidth]{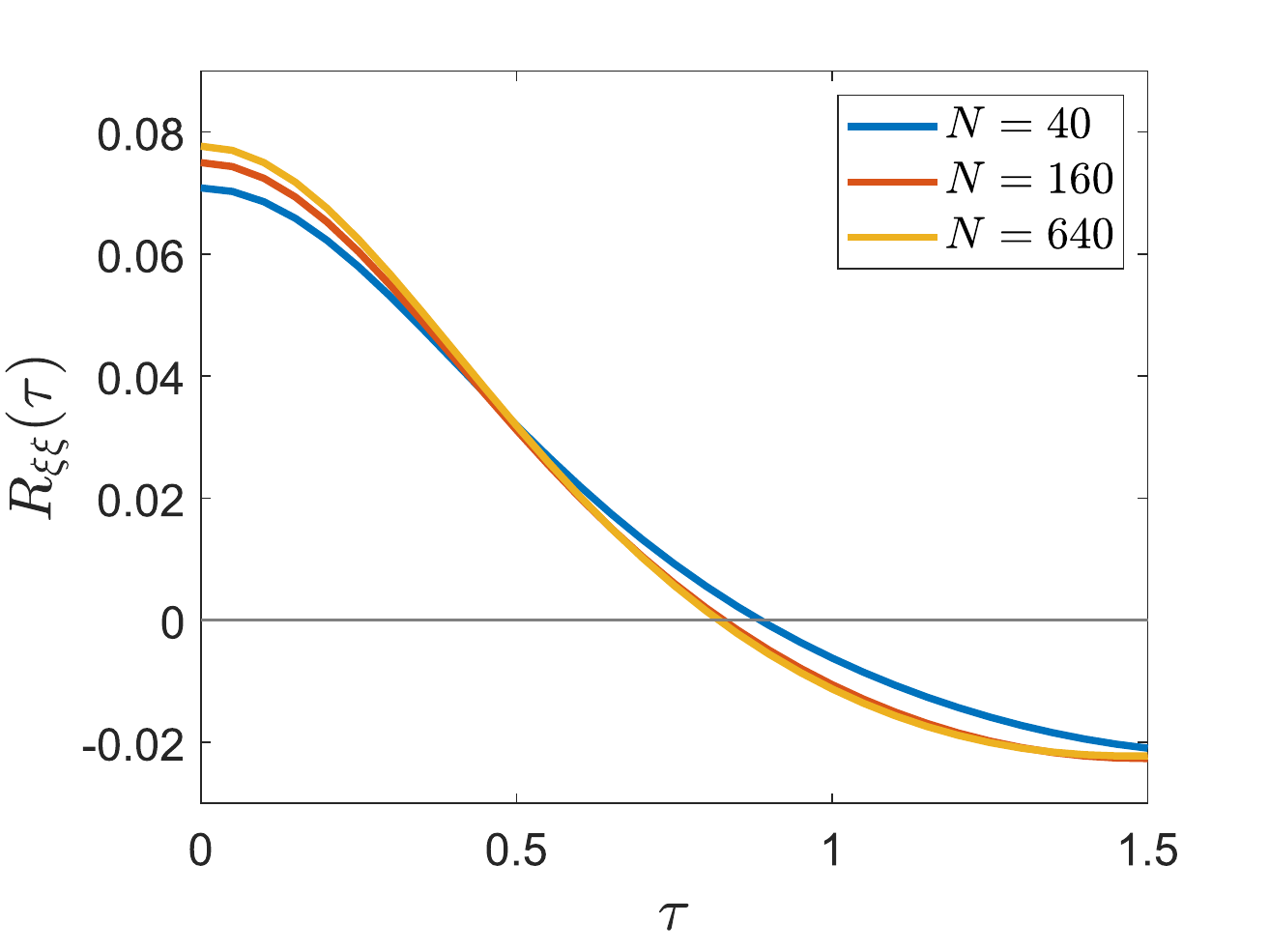}
	\caption{Entry $R_{\xi\xi}(\tau)$ of the covariance function of $Z_{\rm{KS}}=(\xi(t),\zeta(t))$ obtained from a single trajectory of the Kuramoto-Sakaguchi model (\ref{eq:ks}) for different system sizes $N$. We show the covariance function for a large range $\tau\le 40$ as well as a close up for $\tau\le 1.5$.}
	\label{fig:covKM}
\end{figure}

We remark that one cannot expect that the approximate stochastic reduction can model the dynamics accurately on all time scales. 
For example,  the Kuramoto-Sakaguchi model is deterministic and hence the derivatives of the covariance functions $R_{\xi\xi}(\tau)$ and $R_{\zeta\zeta}(\tau)$ are zero at $\tau=0$ \cite{DelSole00} (cf. Figure~\ref{fig:covKM}b). This feature cannot be reproduced by the stochastic Ornstein-Uhlenbeck process which supports covariance functions that are non-differentiable at $\tau=0$; this suggest a lower integration bound $t_{\rm{min}}\neq 0$. We choose here $t_{\rm{min}}= 0.5$ and $t_{\rm{max}}=2.5$. Furthermore, the mismatch between the covariance functions obtained form the full deterministic Kuramoto-Sakaguchi model (\ref{eq:ks}) and those corresponding to the best-fit Ornstein-Uhlenbeck process shown in Figure~\ref{fig:covOU} shows that the effective stochastic dynamics of the fluctuations is only approximately a Gaussian process. We have checked that the accuracy of the fit does not significantly improve when increasing the number of oscillators to $N=2,560$. The discrepancy between the observed covariance function and the best-fit Ornstein-Uhlenbeck process reveals that the underlying Gaussian process is more complex and possibly non-Markovian. However, despite the fact that the OU process is only able to coarsely approximate the covariance function of the actual Gaussian process, we will see in Section~\ref{sec:comp} that the fitted OU-process serves as an easily computable surrogate process for the interaction term of the rogue oscillators, which is able to reliably reproduce the main statistical features of the system.


\subsection{Reduced stochastic model for the entrained oscillators and associated evolution of the order parameter}
Expressing the interaction terms $S(t)$ and $C(t)$ in the Kuramoto-Sakaguchi model (\ref{eq:ks_S_C_sep}) as an OU process with means $\langle S\rangle_t$ and $\langle C\rangle_t$, we obtain the following closed system of evolution equations for the entrained oscillators $\theta_i$ with $i\in \mathcal{C}$ 
\begin{align} 
	d{\theta_i}  &= \big[ \omega_i-\Omega+ \frac{K}{N}\sum_{j\in\mathcal{C}}\sin(\theta_j-\theta_i-\lambda) \,
	\nonumber
	\\
	& \qquad +K\cos \tilde\theta_i \; (\langle S\rangle_t +\frac{1}{\sqrt{N}} \xi_t)  
	\; + \; K\sin \tilde\theta_i  \; (\langle C\rangle_t +\frac{1}{\sqrt{N}}\zeta_t )\; \big]dt ,
\label{eq:ks_sde}
\end{align} 
where $\xi_t$ and $\zeta_t$ are the components of a $2$-dimensional OU process governed by 
\begin{align} 
	d\xi_t &= -\gamma \xi_t \,dt + \upsilon \zeta_t \, dt + \sigma_{11}dB_1 + \sigma_{12} dB_{2} 
	\nonumber 
	\\
	d\zeta_t &= -\gamma \zeta_t\, dt - \upsilon \xi_t \, dt + \sigma_{21}dB_1 + \sigma_{22} dB_{2}.
\label{eq:ks_sde_OU}
\end{align}

We can now establish expressions for the order parameters associated with the stochastic reduced system (\ref{eq:ks_sde})-(\ref{eq:ks_sde_OU}). The order parameter $r_c$ and the mean phase $\psi_c$ can be directly determined from simulations of (\ref{eq:ks_sde})-(\ref{eq:ks_sde_OU}). The full order parameter $r$ can be expressed as
\begin{align}
r e^{i\psi} = \frac{N_c}{N}r_c e^{i\psi_c} +  \frac{N_r}{N}r_r e^{i\psi_r}.
\label{eq:rsde0}
\end{align}
Using the definitions for $S$ and $C$ (\ref{eq:S})-(\ref{eq:C}) which imply
\begin{align*}
(C + i S)e^{-i\lambda} = \frac{N_r}{N} r_r \,e^{i(\psi_r-\psi_c)},
\end{align*}
we can express (\ref{eq:rsde0}) in terms of the complex variable ${\mathcal{Z}}_t=C(t)+i \, S(t)$, where we treat ${\mathcal{Z}}_t$ now as a complex valued random variable, and obtain 
\begin{align}
r(t) = \left| \frac{N_c}{N}r_c(t) + {\mathcal{Z}}_t \, e^{-i\lambda}   \right|.
\label{eq:rsde}
\end{align}
We now perform further approximations to formulate a stochastic evolution equation for the order parameter.  
We separate the mean part and the fluctuations of  ${\mathcal{Z}}_t $ according to 
\begin{align*}
{\mathcal{Z}}_t&=C(t)+i \, S(t)= \langle \mathcal{Z}\rangle + \frac{1}{\sqrt{N}}\mathcal{Z}^\prime_t
\end{align*}
with $\langle \mathcal{Z}\rangle =\langle  C\rangle +i \, \langle  S\rangle$. The random variable ${\mathcal{Z}}^\prime_t=\zeta_t +i\xi_t$ is generated from the OU process (\ref{eq:ks_sde_OU}) via the definition (\ref{eq:xi}). Note that this equation for $r$ is only an approximation as we require $r(t)\le 1$ for all times $t$. 

The order parameters $r_c$ and $r_r$ are both stochastic processes fluctuating around their respective means. However, the variance of the perturbing Ornstein-Uhlenbeck process ${\mathcal{Z}}^\prime _t/\sqrt{N}$, which represents the rogue oscillators, is much larger than that associated with the synchronised cluster. Indeed, we observe numerically a two orders of magnitude larger variance of the rogue oscillators than that of the synchronized oscillators with ${\rm{Var}}(r_c)\approx1.41\times 10^{-5}$ and ${\rm{Var}}(r_r)\approx 6.5\times 10^{-3}$. This suggests that in equation (\ref{eq:rsde}) for the order parameter $r(t)$ the stochasticity of $r_c$ can be neglected to first order against the dominating noise process ${\mathcal{Z}}^\prime _t/\sqrt{N}$, and we can approximate $r_c\approx \bar r_c = {\rm{const}}$. We obtain
\begin{align}
r(t) = \left| \bar r_c^\star + \frac{1}{\sqrt{N}}{\mathcal{Z}}^\prime _t e^{-i\lambda} \right|,
\label{eq:rsdeapprox0}
\end{align}
with complex 
\begin{align}
\bar r_c^\star=\frac{N_c}{N}\bar r_c + \langle \mathcal{Z}\rangle e^{-i\lambda}. 
\end{align}

Assuming that the stochastic perturbations are small we can approximate further and expand up to $\mathcal{O}(1\sqrt{N})$ to obtain
\begin{align}
r(t) = |\bar r_c^\star| + \frac{1}{\sqrt{N}}{\rm{Real}} [\frac{\bar r_c^\star}{|\bar r_c^\star|}e^{-i\lambda}{\mathcal{Z}}_t ^\prime],
\label{eq:rsdeapprox}
\end{align}
which formally is equivalent to the following evolution equation for the order parameter 
\begin{align}
dr = \frac{1}{\sqrt{N}}{\rm{Real}} [\frac{\bar r_c^\star}{|\bar r_c^\star|}e^{-i\lambda} d{\mathcal{Z}}_t^\prime] 
\label{eq:rOA}
\end{align}
with
\begin{align}
 r(0) =|\bar r_c^\star|, \;\;{\mathcal{Z}}^\prime(0) = z^\prime_0.
\label{eq:rOA_IC}
\end{align}
The stochastic process ${\mathcal{Z}}^\prime$ is not necessarily stationary and we allow for non-equilibrium initial conditions $z^\prime_0$ which are not drawn from the stationary distribution of the Ornstein-Uhlenbeck process, corresponding to random initial conditions for the phases of the rogue oscillators. Note that the order parameter is driven by coloured noise in (\ref{eq:rOA}). This is in contrast to Snyder et al \cite{SnyderEtAl21} who postulated Brownian motion as the driving noise. 


\subsection{Dynamic mechanism for the generation of effective stochasticity in the deterministic Kuramoto-Sakaguchi model}
\begin{figure}[] 
\centering
	\includegraphics[width=0.9\linewidth]{{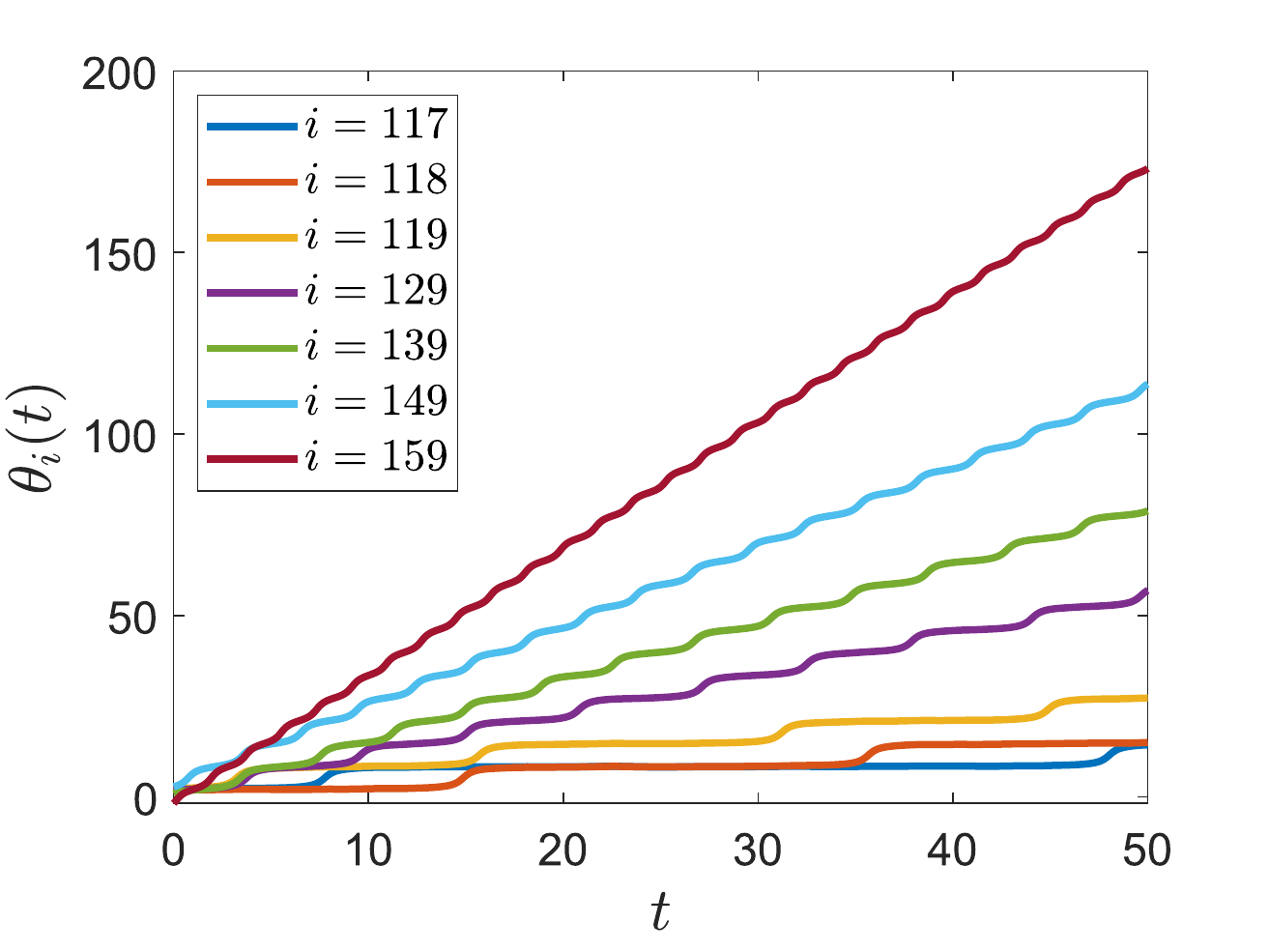}}
	\includegraphics[width=0.9\linewidth]{{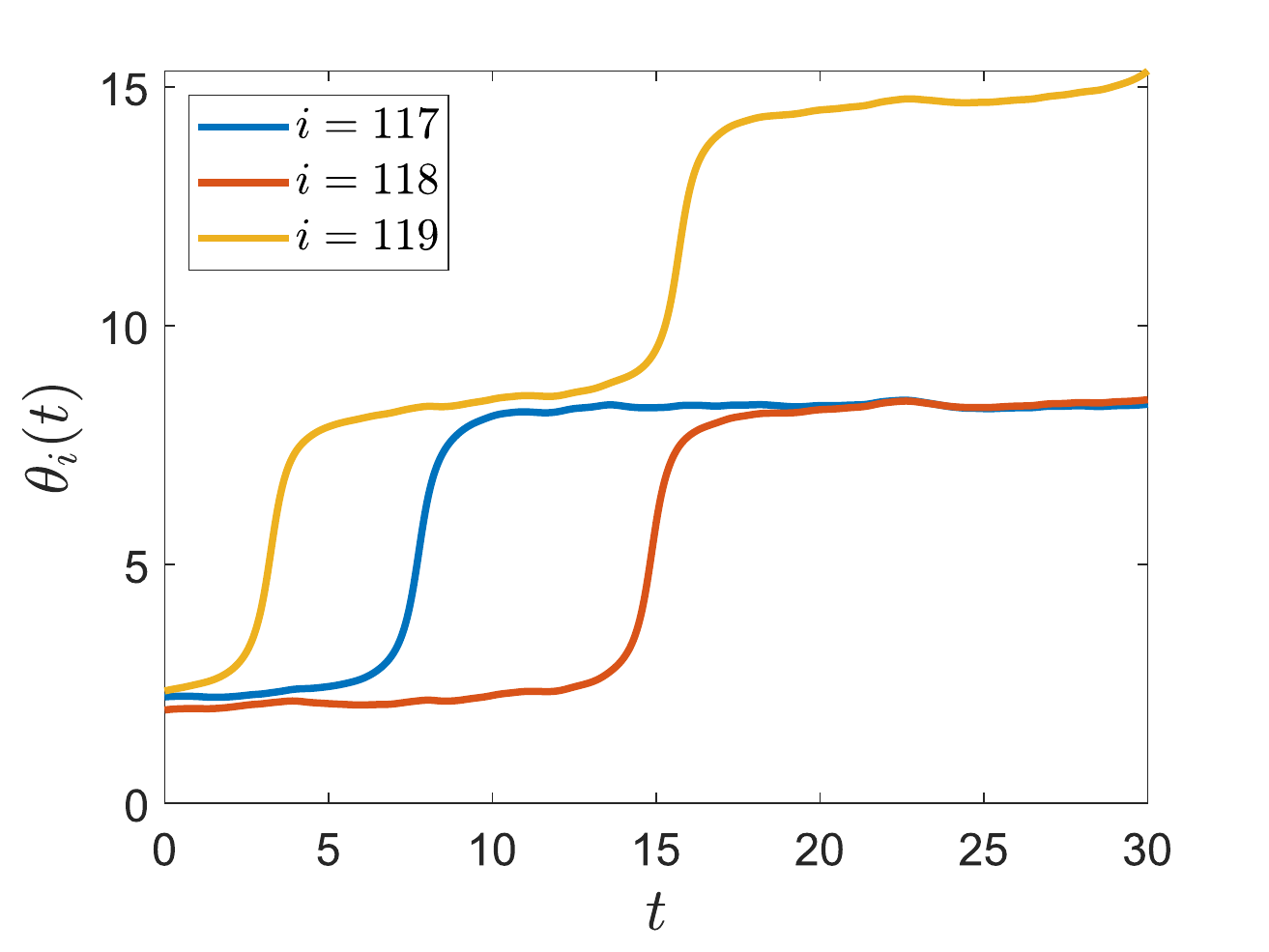}}
		\caption{Top: Typical trajectories of rogue oscillators obtained from a single trajectory of the Kuramoto-Sakaguchi model (\ref{eq:ks}) for $N=160$ including the rogue oscillators with the intrinsic frequency closest to that of an entrained oscillator with index $i=117$ and the rogue oscillator with the largest intrinsic frequency with index $i=159$. Bottom: Zoom into a shorter time window for the rogue oscillators which are closest to the synchronized cluster showing their weakly chaotic nature.}
	\label{fig:rogues}
\end{figure}

The effective stochastic dynamics of $S(t)$ and $C(t)$ are generated by the weak chaoticity of the non-entrained rogue oscillators  \cite{CarluEtAl18,SmithGottwald19}. We show in Figure~\ref{fig:rogues} typical trajectories of rogue oscillators. The dynamics of the weakly chaotic rogue oscillators is characterized by a nearly periodic motion (indeed under the assumption of constant mean-field variables $r$ and $\psi$ the dynamics of each rogue oscillator is approximately governed by the Adler equation (\ref{eq:ksmf})). Note that the rogue oscillators closest to the synchronized cluster evolve slowly, almost aligned with the synchronized cluster, for long periods interrupted by fast slips. We observed that the times between consecutive phase slips are not constant but exhibit significant variance (not shown). In particular, the variance relative to the mean is largest for the slow rogue oscillators closest to the synchronized cluster and decays monotonically for faster rogue oscillators with higher intrinsic frequencies $\omega_i$. The interaction terms (\ref{eq:S}) and (\ref{eq:C}) hence constitute sums of nearly periodic functions of time. 
 It is well known that a sum of many trigonometric functions of linear uncoupled oscillators with uncorrelated random initial conditions approximates a Gaussian processes \cite{Kahane,KupfermanEtAl02,GivonEtAl04}. This suggests that the mechanism for the deterministic generation of diffusive behaviour is less a matter of the time-scale separation between faster chaotic rogue oscillators and slower synchronized oscillators (as would be covered by the theory of homogenization), but instead it is given by weak coupling of the synchronized oscillators to (sufficiently) many uncorrelated rogue oscillators. We remark that the randomness stems here entirely from the choice of the initial conditions of the oscillators. For different initial conditions, different realisations of the noise are generated. The weakly chaotic nature, however, allows one to use highly correlated initial conditions, e.g. $\theta_i={\rm{const}}$ for all $i$; initially correlated phases will decorrelate after a sufficiently long transient period. This is different to the classical trigonometric approximation which relies on the presence of uncorrelated random initial conditions.

 

\begin{figure}[] 
	\includegraphics[width=0.48\linewidth]{{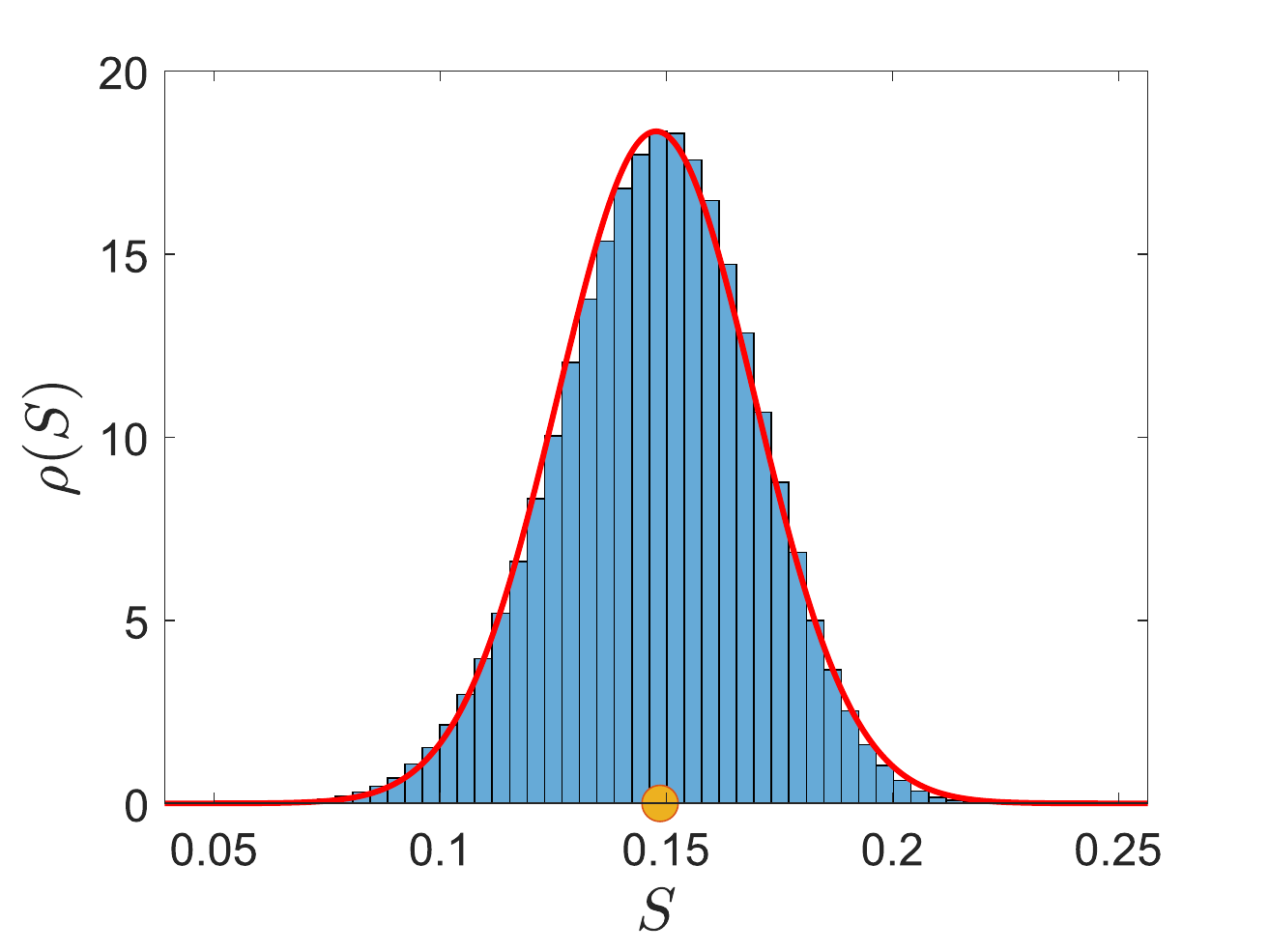}}
	\includegraphics[width=0.48\linewidth]{{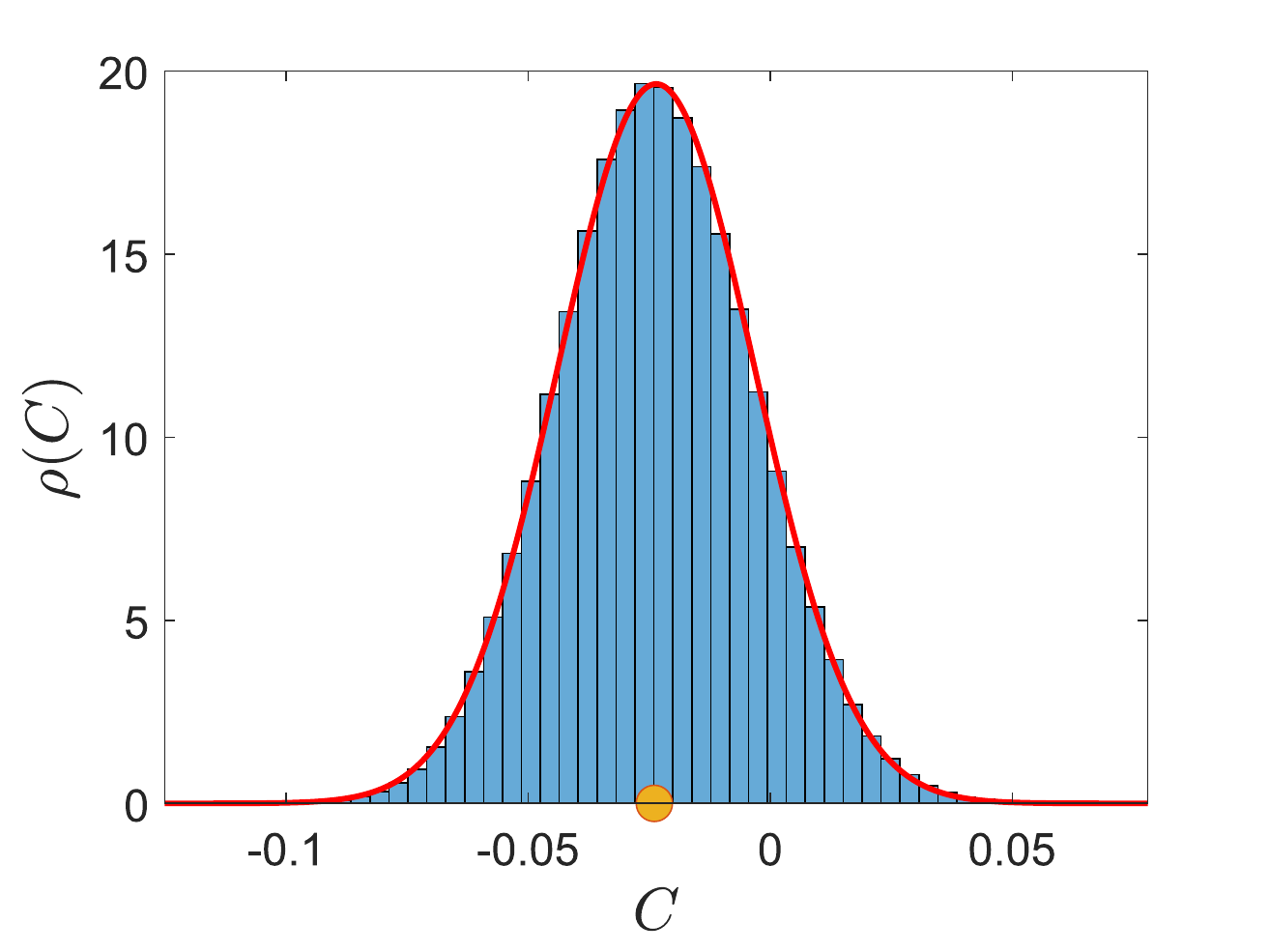}}
	\caption{Empirical histogram of $S(t)$ and $C(t)$ obtained from a single trajectory of the Kuramoto-Sakaguchi model (\ref{eq:ks}) with $N=160$. The continuous curve (online red) shows the corresponding Gaussian best fit. We denote with a filled circle (online orange) the thermodynamic value of $\langle S\rangle_t$ and $\langle C\rangle_t$ as calculated from \eqref{eq:Sbar} and \eqref{eq:Cbar}, respectively.}
	\label{fig:rho_SC}
\end{figure}

\section{Comparison of the stochastic reduced model with the full Kuramoto-Sakaguchi model}
\label{sec:comp}
We now show that the fluctuations of the order parameter as observed in Figure~\ref{fig:rt_with_transient} can be captured by the closed stochastic evolution equation  (\ref{eq:ks_sde})-(\ref{eq:ks_sde_OU}) for the entrained oscillators $\theta_i$ with $i\in \mathcal{C}$. %
Here we present results again for the parameter setting used in Section~\ref{sec:num} with $\lambda = \pi/4$, $g(\omega) \sim \mathcal{N}(0,1)$ and $N = 160$. We begin with a coupling strength $K=3$ as used in Section~\ref{sec:num}.  

The averages $\langle S\rangle_t$ and $\langle C\rangle_t$ in \eqref{eq:ks_sde} can be computed either from long time simulations or from our analytical mean-field theory results \eqref{eq:Sbar} and \eqref{eq:Cbar}. 
The agreement is up to an error of less than $1\%$ (cf. Figure~\ref{fig:rho_SC}). We use in the following the numerically obtained values. The parameters of the Ornstein-Uhlenbeck process \eqref{eq:ks_sde_OU} were determined by the fitting of the covariance functions according to \eqref{eq:LSQ} which we recall here as $\gamma=0.727, \upsilon=1.73$ and $\sigma_{11}=0.374, \sigma_{12}=\sigma_{21}=0.0090, \sigma_{22}=0.271$. We remark that in the thermodynamic limit $N\to\infty$ the effect of the stochastic fluctuations $\xi_t$ and $\zeta_t$ in (\ref{eq:ks_sde}) decays and we recover the deterministic evolution equation derived in \cite{YueEtAl20}. We simulate the system (\ref{eq:ks_sde})-(\ref{eq:ks_sde_OU}) using an Euler-Maruyama integrator with a time step of $dt=0.01$. 

\begin{figure}
\centering
         \includegraphics[width=0.9 \linewidth]{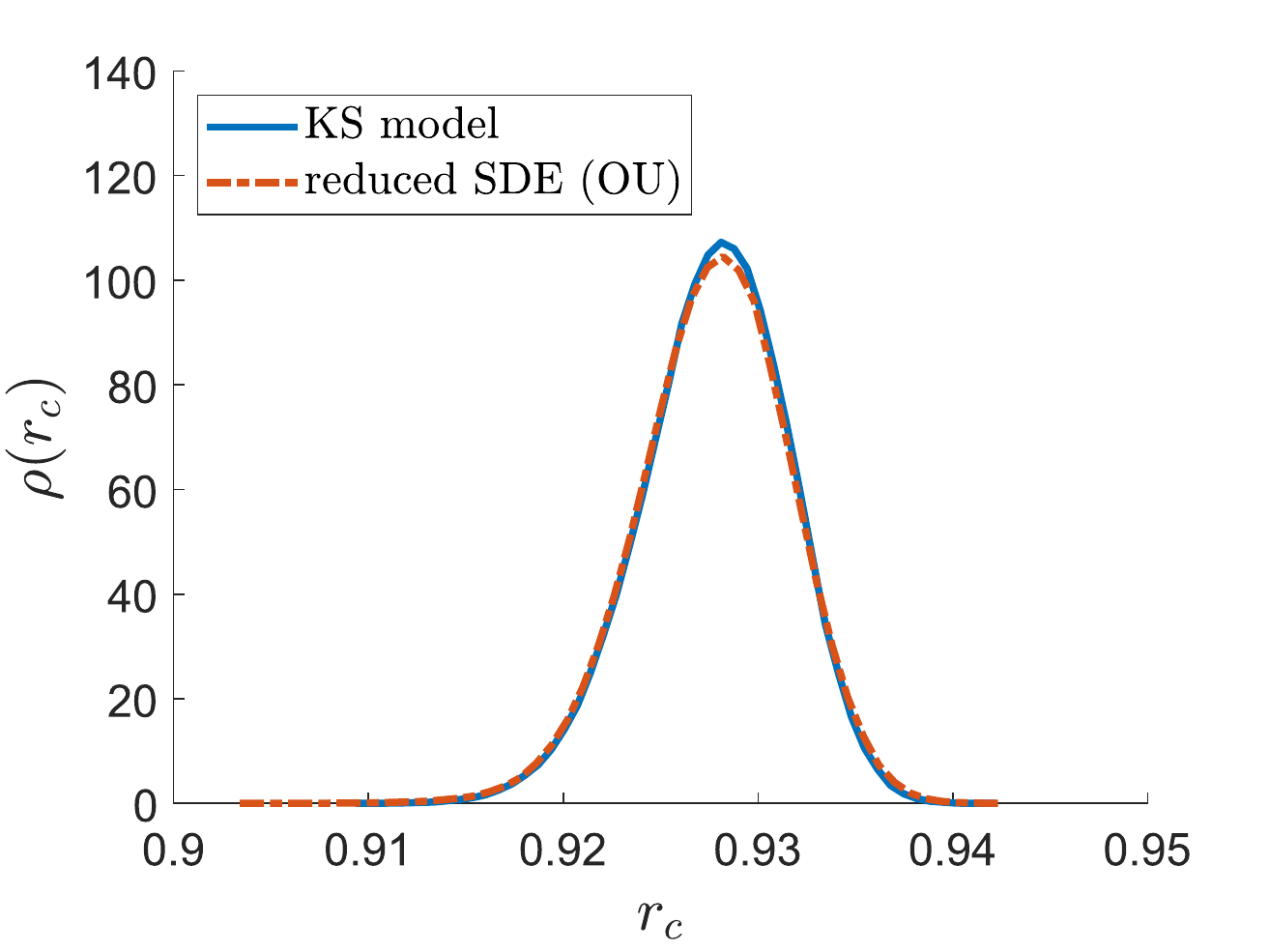}
         \includegraphics[width=0.9 \linewidth]{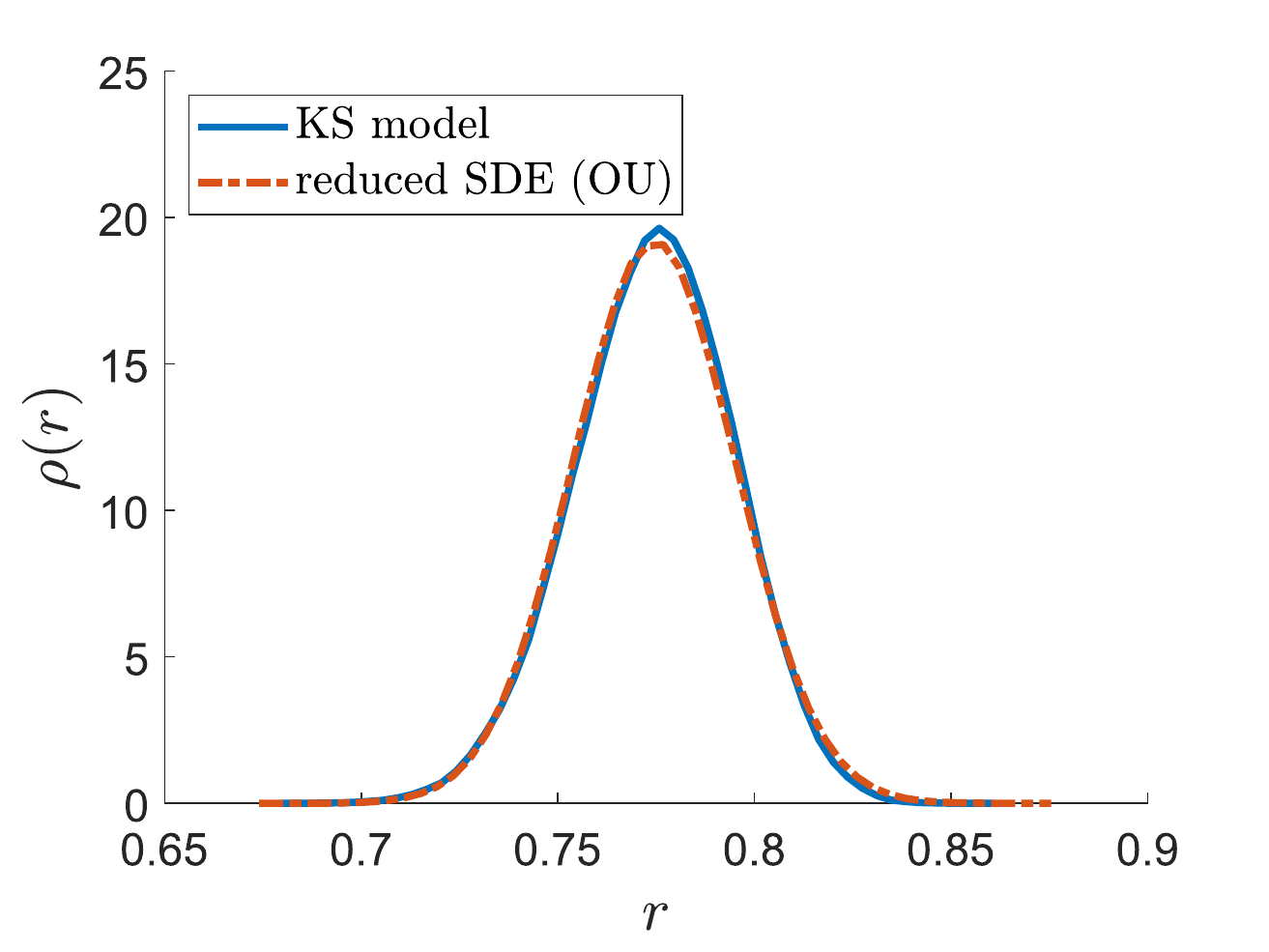}
	\caption{Comparison of the empirical histograms for $r_c$ (left) and $r$ (right) obtained from a single trajectory of the Kuramoto-Sakaguchi model (\ref{eq:ks}) for $N=160$ and from the dynamics of the reduced stochastic model (\ref{eq:ks_sde}) driven by an OU process. 
	All other parameters are the same as in Fig.~\ref{fig:snapshot}.}
	\label{fig:rho_r_rc}
\end{figure}

Figure~\ref{fig:rho_r_rc} shows the empirical histograms of the order parameter $r_c$ pertaining to the synchronized oscillators as well as the total order parameter $r$ for all oscillators when simulated using the full Kuramoto-Sakaguchi model (\ref{eq:ks}) 
and when estimated by the reduced stochastic system (\ref{eq:ks_sde})-(\ref{eq:ks_sde_OU}) using (\ref{eq:rc}) and (\ref{eq:rsde}) for the associated order parameters; we remark that results using the approximation (\ref{eq:rsdeapprox0}) for the associated order parameter lead to results indistinguishable by eye. We further show the histogram of the order parameter $r_c$ which only takes into account the entrained synchronized oscillators. It is seen that our stochastic reduction captures the observed fluctuations for both $r_c(t)$ and $r(t)$ very well. We remark that the histograms are not distinguishable by eye if the order parameter is calculated by an actual time trajectory of $r_c(t)$ or by its constant mean $\bar{r}_c$ (cf. (\ref{eq:rsde}) and (\ref{eq:rsdeapprox0})). 

\begin{figure}
\centering
         \includegraphics[width=0.9 \linewidth]{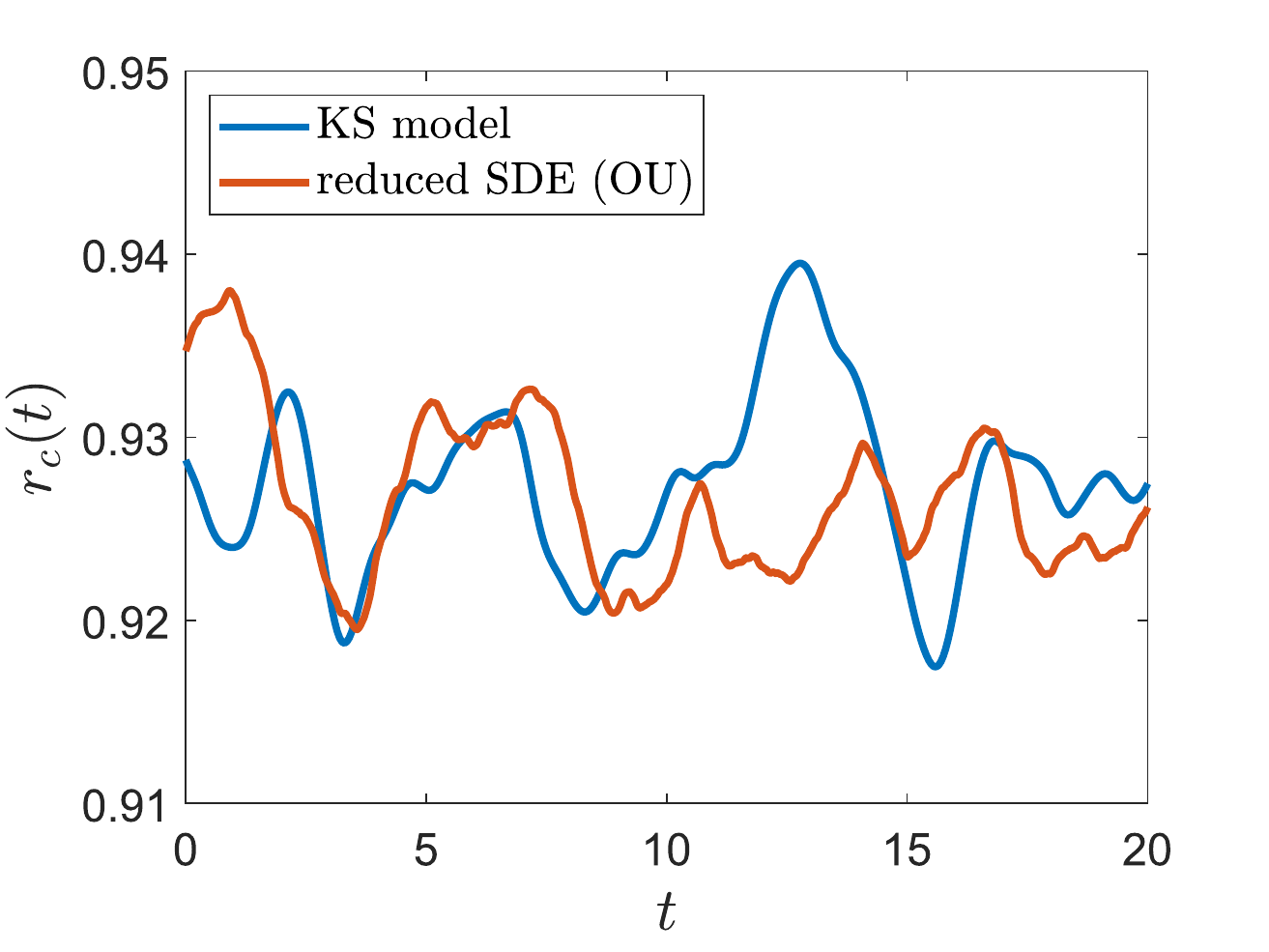}
	\caption{Temporal evolution of the order parameter $r_c(t)$. Shown are results obtained from the full Kuramoto-Sakaguchi model (\ref{eq:ks}) (online blue), the reduced stochastic model (\ref{eq:ks_sde})-(\ref{eq:ks_sde_OU}) driven by an OU process (online red).
	All other parameters are the same as in Fig.~\ref{fig:snapshot}.}
	\label{fig:rcomp}
\end{figure}

Long-time solutions of the reduced stochastic equation for the order parameter (\ref{eq:rOA}) generate empirical histograms undistinguishable by eye from those presented in Figure~\ref{fig:rho_r_rc}. 
In Figure~\ref{fig:rcomp} we show the time evolution of the order parameter $r_c(t)$ computed from simulations of the full Kuramoto-Sakaguchi model (cf. Figure~\ref{fig:rt_with_transient}) and of our reduced stochastic system (\ref{eq:ks_sde})-(\ref{eq:ks_sde_OU}). It is seen that the qualitative smooth character of the fluctuations observed in the deterministic Kuramoto-Sakaguchi model (\ref{eq:ks}) is reproduced by our stochastic system which is driven by coloured OU noise.\\

\begin{figure}
\centering
         \includegraphics[width=0.49 \linewidth]{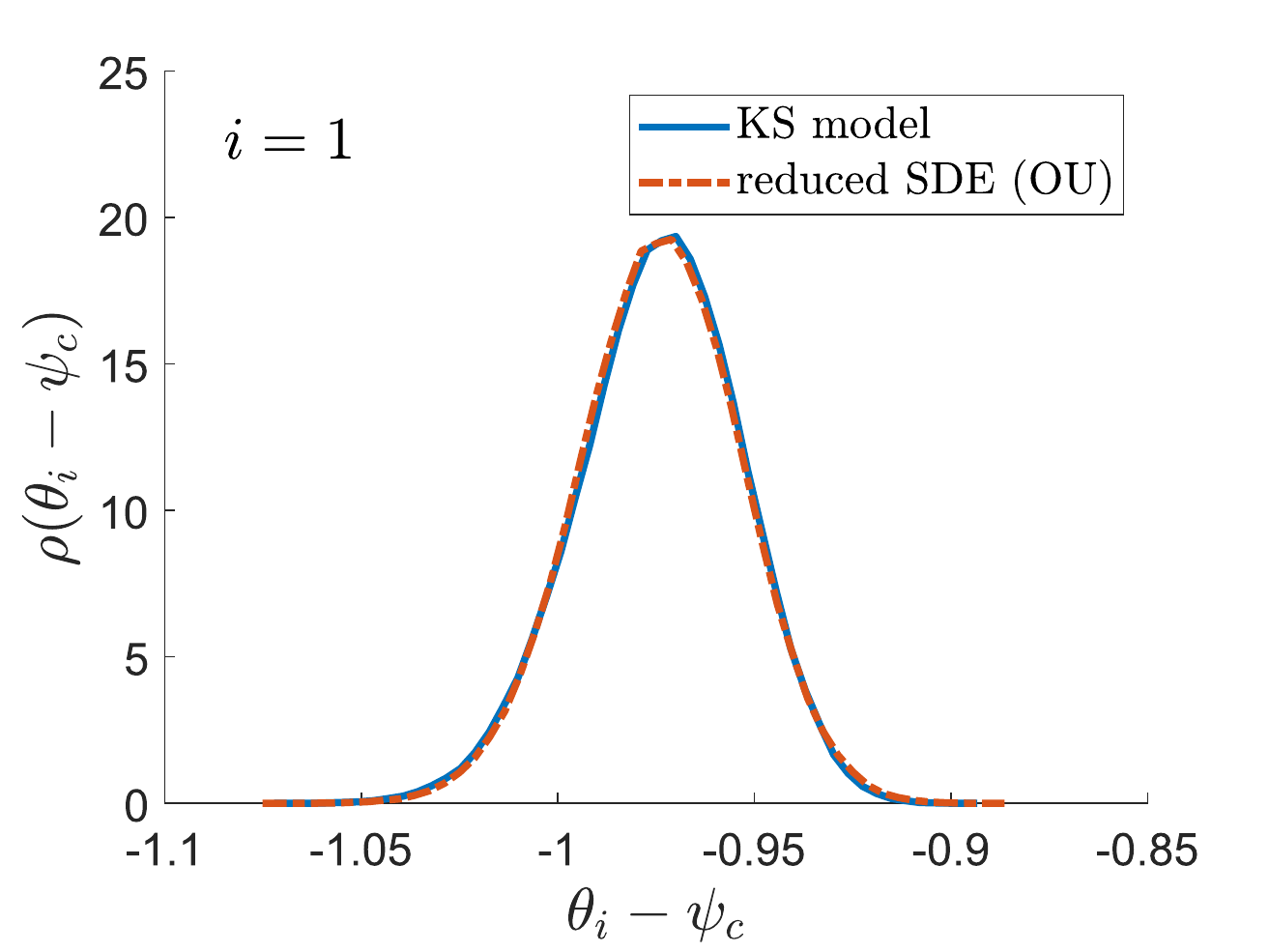}\\
         \includegraphics[width=0.49 \linewidth]{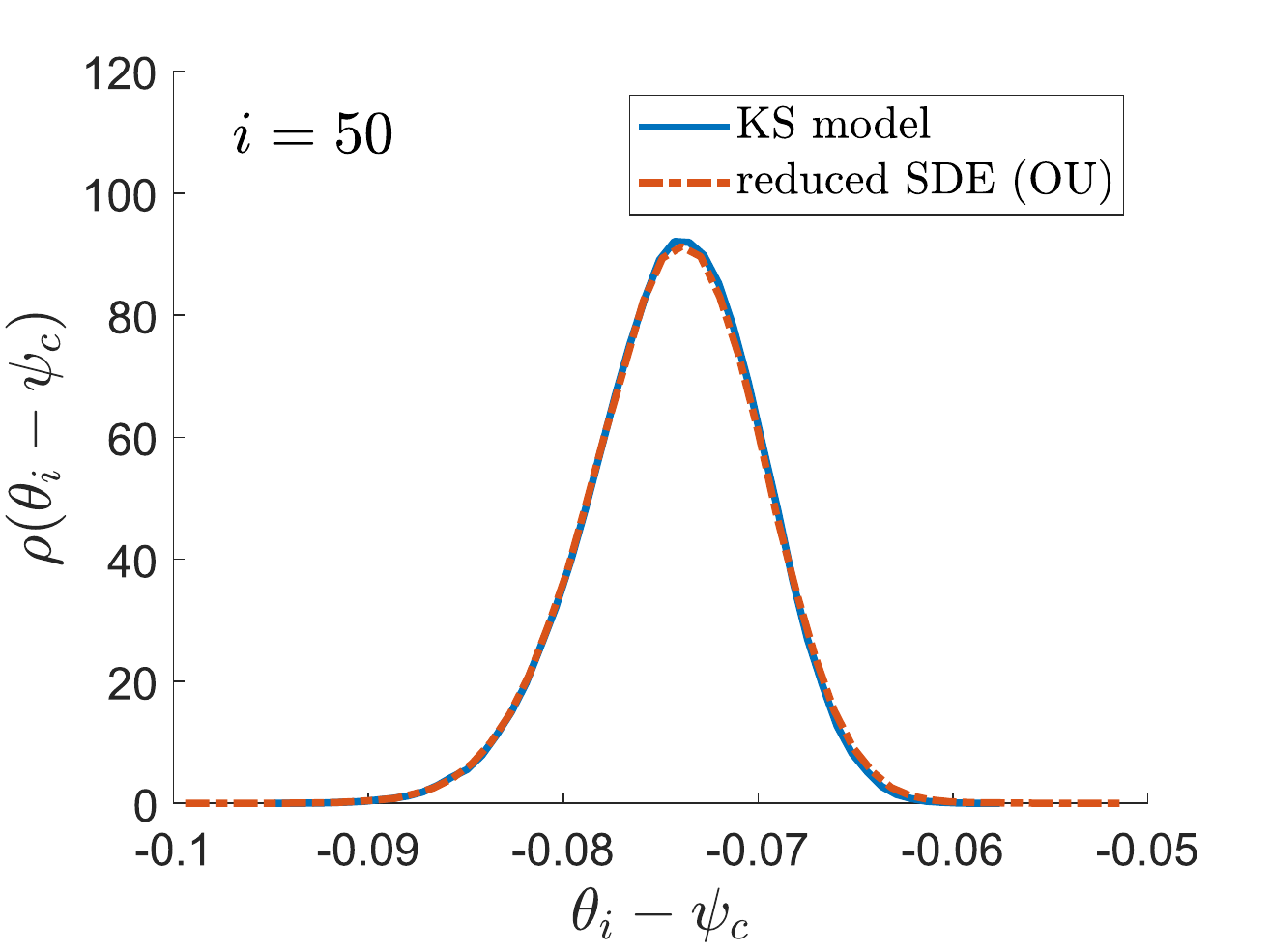}
         \includegraphics[width=0.49 \linewidth]{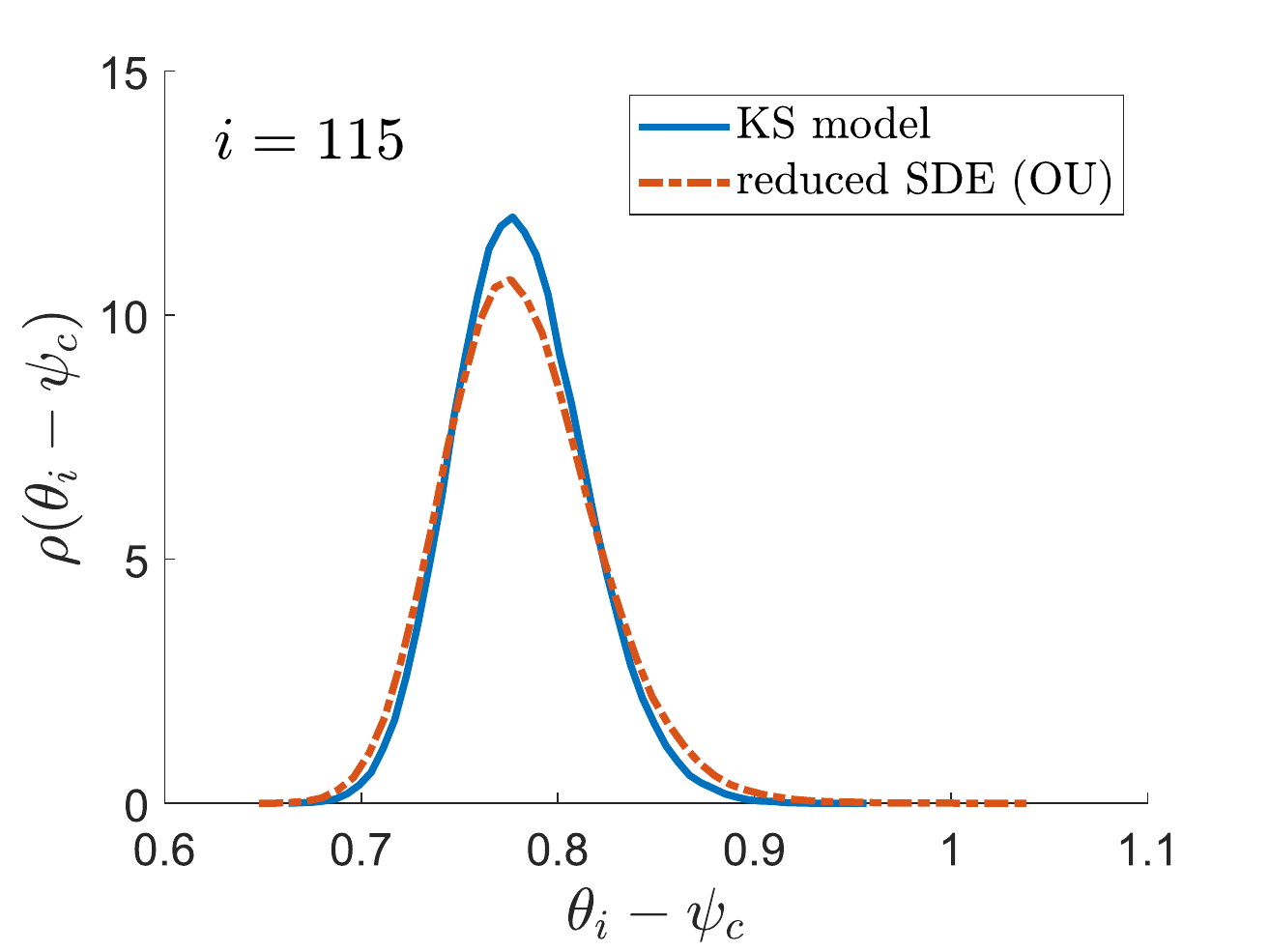}
	\caption{Empirical density of $\theta_i-\psi_c$ for the entrained oscillators with indices $i=1$, $i=50$ and $i=115$. Shown are results for the full Kuramoto-Sakaguchi model (\ref{eq:ks}) for $N=160$ and for the reduced stochastic model (\ref{eq:ks_sde})-(\ref{eq:ks_sde_OU}). All other parameters are the same as in Fig.~\ref{fig:snapshot}.}
	\label{fig:rho_theta}
\end{figure}

\begin{figure}
\centering
         \includegraphics[width=0.9 \linewidth]{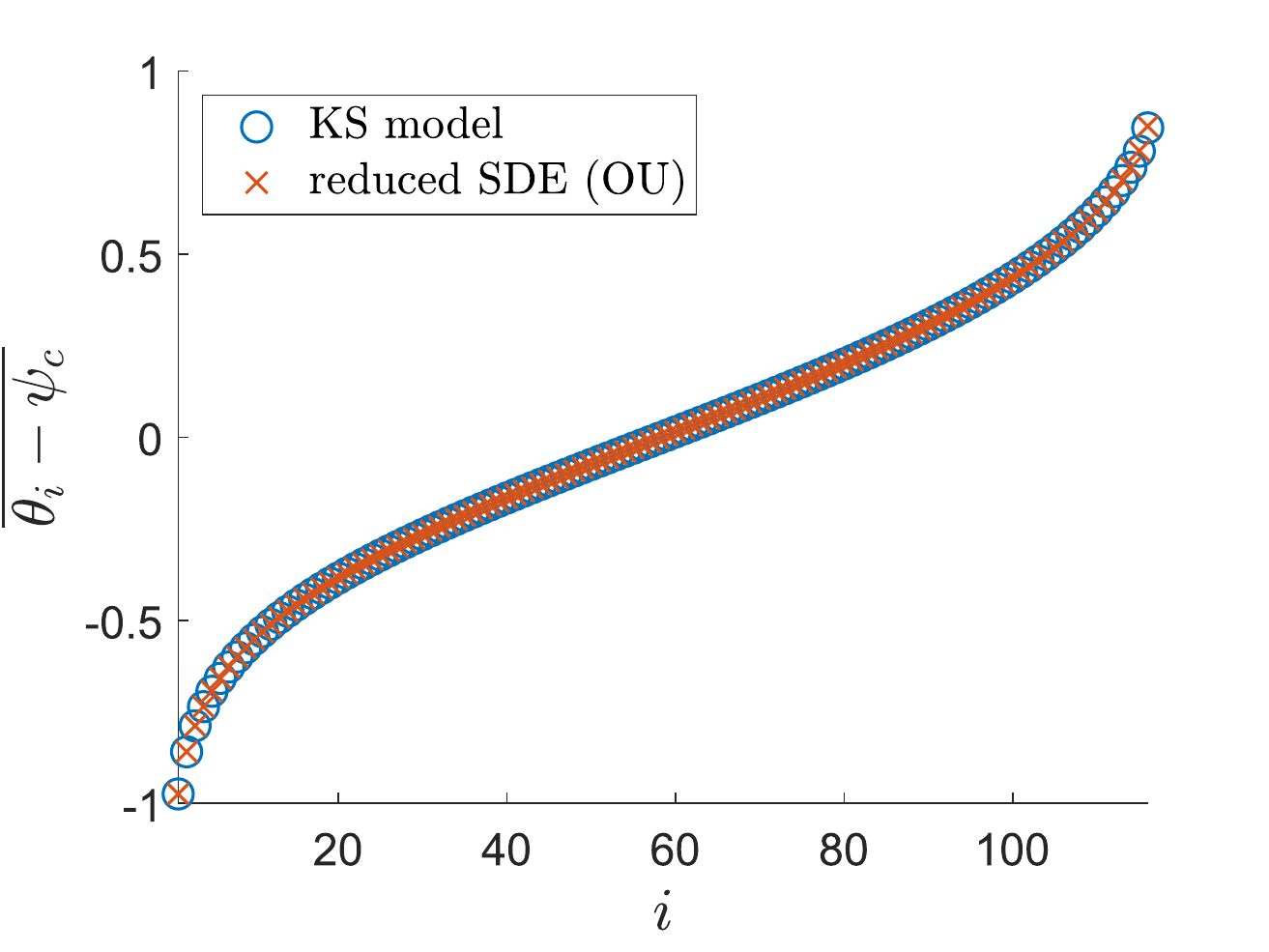}
         \includegraphics[width=0.9 \linewidth]{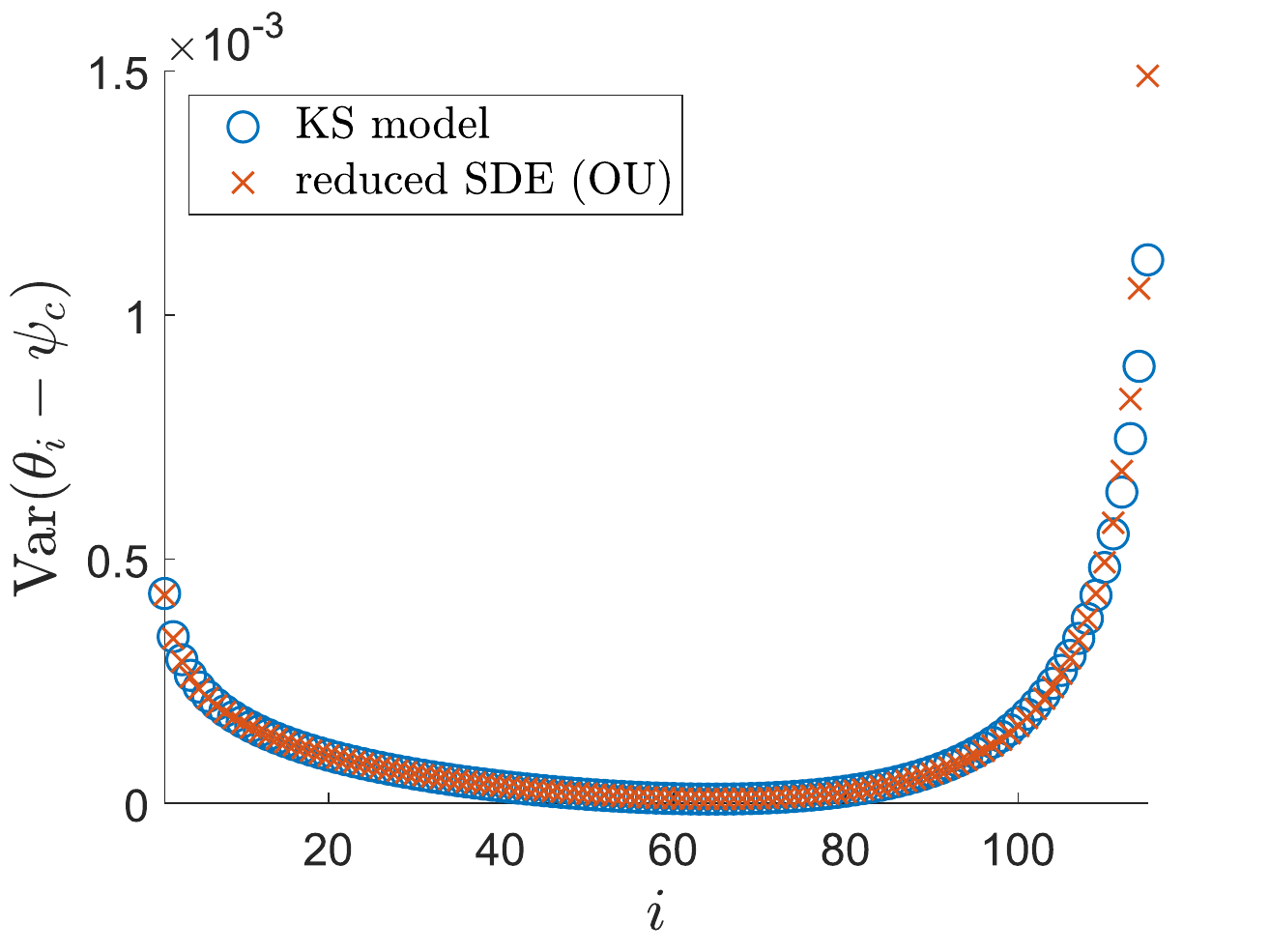}
	\caption{Mean (left) and variance (right)) of the fluctuations of the entrained synchronized oscillators around their mean phase, $\theta_i-\psi_c$ for all $i\in\mathcal{C}$.  Shown are results for the full Kuramoto-Sakaguchi model (\ref{eq:ks}) for $N=160$ and for the reduced stochastic model (\ref{eq:ks_sde})-(\ref{eq:ks_sde_OU}). All other parameters are the same as in Fig.~\ref{fig:snapshot}.}
	\label{fig:theta_fluct}
\end{figure}

To further probe the ability of the reduced stochastic model (\ref{eq:ks_sde})-(\ref{eq:ks_sde_OU}) to capture the collective dynamics of the entrained synchronized oscillators we show results for the fluctuations of the entrained oscillators around the mean phase of the synchronized cluster, $\theta_i-\psi_c$ for $i\in\mathcal{C}$. These fluctuations are Gaussian for all entrained oscillators as shown in Figure~\ref{fig:rho_theta}. It is seen that our SDE very well describes entrained oscillators, such as those with indices $i=1$ and $i=50$, but the degree of approximation is less good for those entrained oscillators at the edge of the cluster with index $i=115$. In Figure~\ref{fig:theta_fluct} we show the mean and the variance of $\theta_i-\psi_c$ for all $i\in\mathcal{C}$ estimated from a long simulation of the full Kuramoto-Sakaguchi model (\ref{eq:ks}) and of the reduced stochastic model (\ref{eq:ks_sde})-(\ref{eq:ks_sde_OU}). It is seen that the mean is very well recovered by the reduced stochastic model. The variance is very well captured for oscillators with an index $i$ that is sufficiently small to ensure that their intrinsic frequency is not too close to that of the closest rogue oscillator. For the simulations we show in Figure~\ref{fig:theta_fluct} the index of the first rogue oscillator is $i=117$ for the full Kuramoto-Sakaguchi model. The entrained oscillator with index $i=116$, i.e the oscillator on the edge of the synchronized cluster, is fully entrained in the Kuramoto-Sakaguchi model. However, for the approximate stochastic model (\ref{eq:ks_sde})-(\ref{eq:ks_sde_OU}) the oscillator with index $i=116$ experiences rare fast random slips between long periods of entrainment (of the order of $1,000$ time units on average). The rare and fast slips do not affect the mean which is still close to the mean corresponding to the Kuramoto-Sakaguchi model, but the variance is much higher with a value of $0.012$ (not shown in Figure~\ref{fig:theta_fluct} where we only show oscillators with index $i\le 115$). \\  

\begin{figure}
\centering
         \includegraphics[width=0.9 \linewidth]{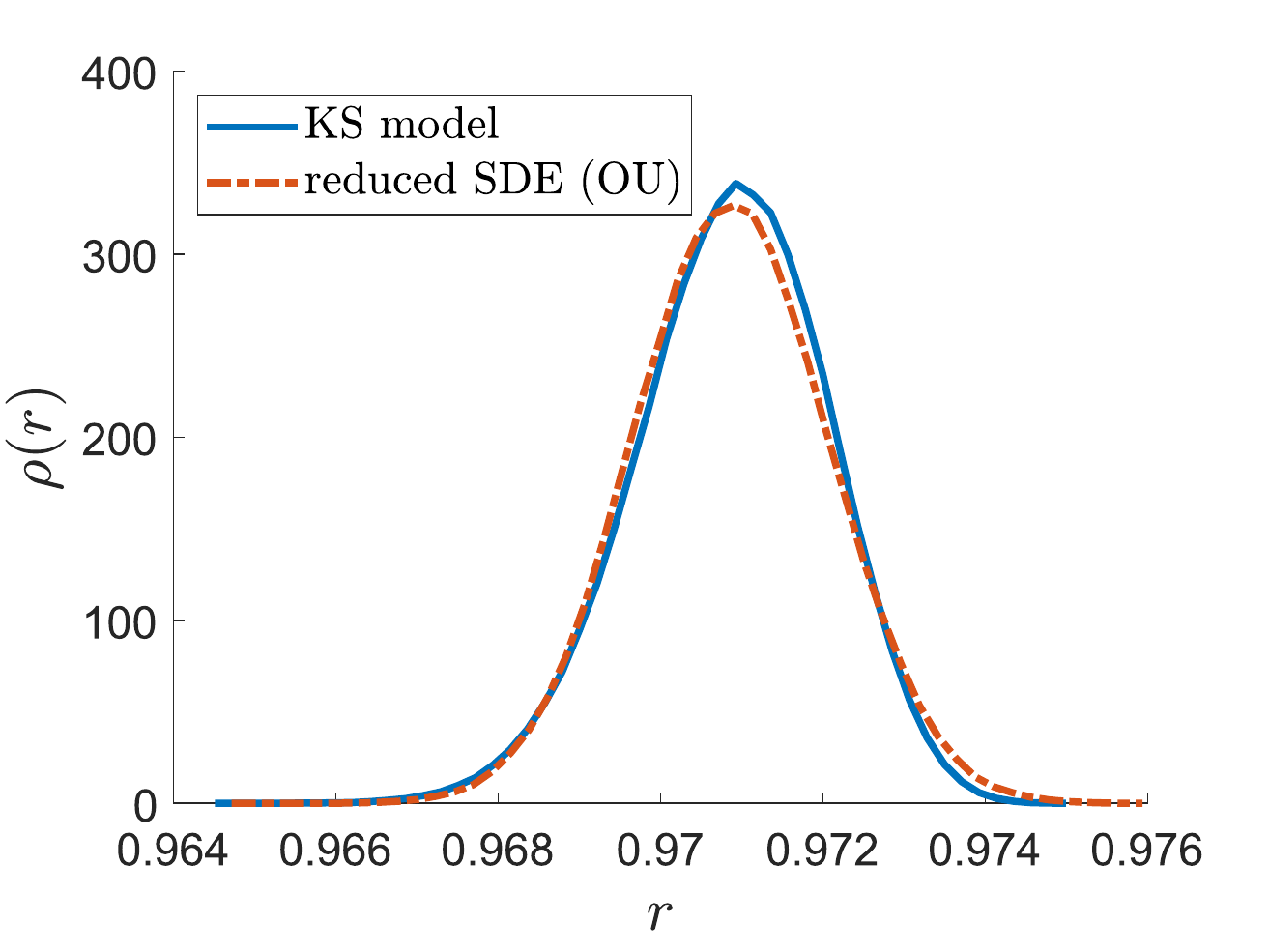}
       \includegraphics[width=0.9 \linewidth]{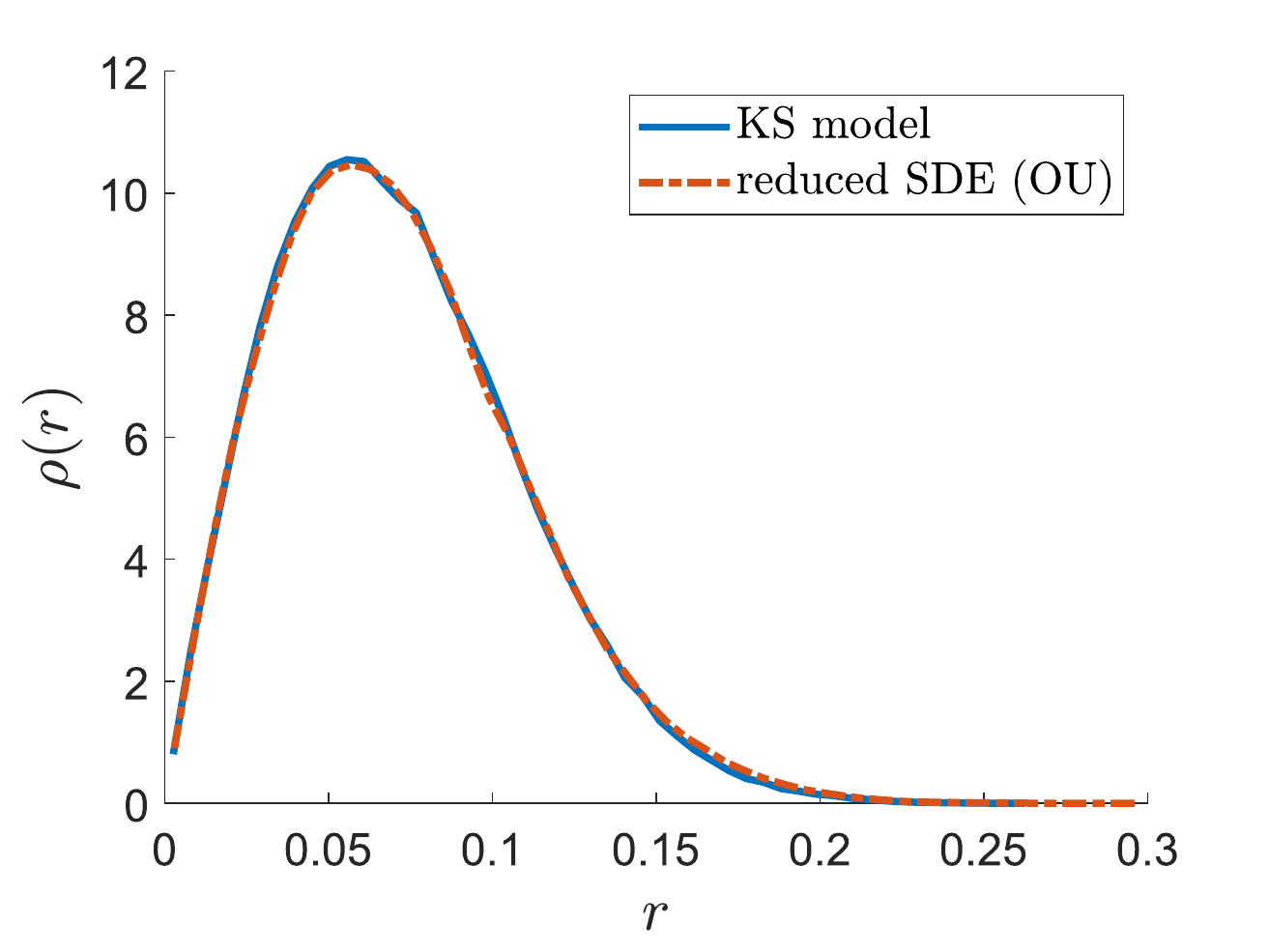}
	\caption{Comparison of the empirical histograms for the order parameter $r$ obtained from a single trajectory of the Kuramoto-Sakaguchi model (\ref{eq:ks}) and from the dynamics of the reduced stochastic model (\ref{eq:ks_sde}) driven by an OU process. Top: For coupling strength $K=7$ with $N=2,560$. The best-fit parameters for the Ornstein-Uhlenbeck process (\ref{eq:ks_sde_OU}) are $\gamma=1.2947,\upsilon=2.7992,\sigma_{11}=0.0980, \sigma_{12} = \sigma_{21}=0.0007$ and $\sigma_{22}=0.0955$. The order parameter of the reduced stochastic model (\ref{eq:ks_sde}) is calculated using (\ref{eq:rsde}). Bottom: For the subcritical coupling strength $K=1.25$ for which there are no synchronized clusters, with $N=160$. Here the reduced stochastic model (\ref{eq:ks_sde}) is for a single oscillator with index $i=1$, with $N_r=N-1=159$ and $\psi_c=\theta_1$. The best-fit parameters for the Ornstein-Uhlenbeck process (\ref{eq:ks_sde_OU}) are $\gamma = 0.5454$, $\upsilon = 1.8682$, $\sigma_{11}= 0.7513$, $\sigma_{12}=\sigma_{21} = -0.0019$ and $\sigma_{22} =  0.7496$. The order parameter of the reduced stochastic model (\ref{eq:ks_sde}) is calculated using (\ref{eq:rsde}).}
	\label{fig:r_subcritical}
\end{figure}

We have so far reported only on a single value $K=3$ of the coupling strength. It is pertinent to mention that our approach is not restricted to any value of the coupling strength but only to the number of rogue oscillators present. For each value of the coupling strength, however, the corresponding parameters of the approximating Ornstein-Uhlenbeck process will differ and will need to be determined. For $K=3$ we found that the number of rogue oscillators $N_r=43$ out of a total of $N=160$ oscillators was sufficient for the fluctuations to be described as a Gaussian process. The number of rogue oscillators decreases with increasing coupling strength, see Figure~\ref{fig:ks_diff_K}. For an increased coupling strength with $K=7$ there are only $N_r=2$ rogue oscillators for $N=160$ which is insufficient for the Gaussian process approximation. However, increasing the total number of oscillators to $N=2,560$ increases the number of rogue oscillators to $N_r=46$, and we observe again that the statistics of the order parameter is well recovered by the Gaussian process approximation as shown in Figure~\ref{fig:r_subcritical}. Conversely, for coupling strengths below the critical coupling strength $K_c$ one can use the Gaussian approximation for any of the $N$ oscillators with the remaining $N_r=N-1$ oscillators contributing to an additive noise for the single oscillator with index $i=i^\star$ and $\psi_c=\theta_{i^\star}$. Figure~\ref{fig:r_subcritical} shows that indeed the order parameter $r$ is well approximated by the Gaussian process approximation for the sub-critical case $K=1.25$.\\

The numerical results reported above all used the equiprobable draws of the intrinsic frequencies described in Section~\ref{sec:num}. For random draws the number of rogue oscillators $N_r$ fluctuates across realisations, and the statistical features of $S$ and $C$ will vary. We checked that with random draws of the intrinsic frequencies the distributions of $S$ and $C$ are still near-Gaussian and $S$ and $C$ can be effectively approximated by a Gaussian process for most realisations. For random intrinsic frequency realisations which give rise to the formation of smaller interacting clusters, the effective dynamics would need to involve multiple clusters and becomes much more involved. The fluctuations decrease with increasing number of oscillators $N$ approaching the results of the equiprobable draws presented here. 

\section{Discussion}
\label{sec:disc}

We considered  
the effect of non-entrained rogue oscillators on the collective behaviour of the entrained synchronized oscillators in the Kuramoto-Sakaguchi model. We established an average effect, in the spirit of a law of large numbers, and then considered fluctuations around the mean effect akin to the central limit theorem. We presented results from numerical simulations that suggest that the fluctuations can be approximated by a Gaussian process with a stationary density and fluctuations which decay as $1/\sqrt{N}$. Hence for fluctuations to have a significant effect one needs the number of rogue oscillators to be sufficiently large to allow for a central limit theorem and sufficiently small to have a nonnegligible variance. Gaussian processes are determined entirely by their mean and their covariance function. We used a two-dimensional Ornstein-Uhlenbeck process as a surrogate stochastic Gaussian process by fitting the covariance function. 
This enabled us to replace the interaction term involving the rogue oscillators by this Ornstein Uhlenbeck process, and to formulate a closed equation for the entrained synchronized oscillators only, which are driven by this complex Ornstein-Uhlenbeck process. The reduced system of stochastic differential equations showed remarkable capability to reproduce the statistical behaviour of the entrained oscillators such as the probability density function of the order parameter. It is pertinent to mention that the order parameter is driven by coloured Ornstein-Uhlenbeck noise and not, as previously claimed on heuristic grounds, by Brownian motion. The OU noise was fitted by directly using information of the unresolved dynamics (i.e. the statistical behaviour of $S$ and $C$) rather than by blind fitting of the resolved synchronized behaviour, as is usually done when imposing coarse grained model.\\

Our numerical experiments were restricted to normally distributed intrinsic frequencies. There are in principle no restrictions to the nature of the frequency distribution $g(\omega)$. For unimodal frequency distributions and also for uniform frequency distributions with $\lambda \neq 0$ there is a range in coupling strengths for which the dynamics is characterized by  the interaction of a single partially synchronized cluster with a collection of non-entrained rogue oscillators as discussed here. It would be interesting to see how finite-size induced noise can effect the interaction of several partially synchronized clusters. This would be possible for multimodal intrinsic frequency distributions.\\ 

In future work it would be interesting to investigate in how far the stochastic model reduction is able to characterize the fluctuations of the order parameter at the phase transition as seen in Figure~\ref{fig:ks_diff_K}b. Since the parameters of the best-fit OU process typically differ for different values of the coupling strength $K$ this may prove to be numerically costly.\\

On a theoretical level, it would be interesting to show the convergence to an effective stochastic equation for the synchronized oscillators more rigorously and/or find explicit expressions for the parameters of the reduced SDE. This would require to extend the method of trigonometric approximations to an approximation of Gaussian processes by a sum of solutions of the Adler equation \eqref{eq:ksmf}. This could then allow for a computationally feasible investigation of the fluctuations of the order parameter at the phase transition, mentioned above.\\

The stochastic reduced equation can now be further reduced, for example via the method of collective coordinates \cite{Gottwald15,HancockGottwald18,SmithGottwald19,YueEtAl20,SmithGottwald20}. In particular, one can employ the methods developed for the reduction of a stochastic Kuramoto model \cite{Gottwald17} to the proposed stochastic system here for the Kuramoto-Sakaguchi model. This has the potential to capture finite size effects which cannot be captured via standard mean-field theories, in particular, the diffusivity of the mean phase $\psi$ and $\psi_c$. 


\section*{Acknowledgments}
We thank Lauren Smith for numerous valuable discussions. GAG acknowledges support from the Australian Research Council (Grant No. DP180101991).

\section*{Data availability}
Code is available at \url{https://github.com/wenqiyue/ks_stoch_approx}.

\appendix
\section{Covariance function of the $2$-dimensional Ornstein-Uhlenbeck process}
\label{sec:app}
We write the $2$-dimensional OU process \eqref{eq:OU} as
\begin{align} 
	dz_t 
		&= L z_t \, dt + \Sigma \, dB_t,
\label{eq:app1}
\end{align}
where $L = -\Gamma+\Upsilon$. Formally, the solution of this SDE with initial condition $z_0$ can be written as
\begin{align*} 
	z_t = e^{Lt}z_0 + \int_{0}^{t}e^{L(t-u)}\Sigma \, dB_u.
\end{align*}
Defining for the mean-zero process $z_t$
\begin{align*} 
	R(t,t+\tau) &= \E[z_tz^\tr_{t+\tau}] \\
	&=e^{Lt} \E\left[z_0 z_0^\tr\right]e^{L^\tr(t+\tau)} + \int_0^t e^{L(t-u)}\Sigma\Sigma^\tr e^{L^\tr(t+\tau-u)}du,
\end{align*}
where the first expectation is taken with respect to Brownian motion paths and initial conditions and the second expectation with respect to initial conditions, the covariance function is obtained as 
\begin{align*} 
	R(\tau) = \lim\limits_{t\rightarrow\infty}R(t,t+\tau) = \lim\limits_{t\rightarrow\infty} \int_0^t e^{L(t-u)}\Sigma\Sigma^\tr e^{L^\tr(t+\tau-u)}du.
\end{align*}
Restricting to a diagonal matrix $\Gamma=\gamma\, \I$ and a skew-symmetric rotation matrix $\Upsilon$ with entries $\upsilon_{12} = -\upsilon_{21}= \upsilon$ and $\upsilon_{11}=\upsilon_{22}=0$, and to a symmetric diffusion matrix $\Sigma$ with entries $\sigma_{ij}$ for $i,j=1,2$, the $2$-dimensional mean-zero OU process \eqref{eq:app1} is written as \eqref{eq:ks_sde_OU}, which we recall here as
\begin{align*} 
	\begin{pmatrix} d\xi_t\\ d\zeta_t\end{pmatrix} = \pmat{-\gamma &\upsilon \\ -\upsilon &-\gamma} \pmat{\xi_t \\ \zeta_t} dt + \pmat{\sigma_{11} &\sigma_{12} \\ \sigma_{12} &\sigma_{22}} \pmat{dB_{1,t}\\dB_{2,t}}.
\end{align*}
The covariance function for this process can be explicitly expressed as
\begin{align*} 
	R^{(OU)}(\tau) = \frac{1}{4\gamma(\gamma^2+\upsilon^2)}e^{-\gamma\tau}\hat{R}(\tau)
\end{align*}
with
\begin{align*} 
	\hat{R}(\tau) = \pmat{\hat{R}_{\xi \xi}(\tau) &\hat{R}_{\xi \zeta}(\tau) \\ \hat{R}_{\zeta \xi}(\tau) &\hat{R}_{\zeta \zeta}(\tau)}. 
\end{align*}
Defining 
\begin{align*} 
	\pmat{a_1&a_2\\a_2 &a_3} = \pmat{\sigma_{11} &\sigma_{12}\\ \sigma_{12} &\sigma_{22}}\pmat{\sigma_{11} &\sigma_{12}\\ \sigma_{12} &\sigma_{22}}^\tr,
\end{align*}
i.e.
\begin{align*} 
	a_1 = \sigma_{11}^2+\sigma_{12}^2 \hspace{20pt} a_2 = \sigma_{11}\sigma_{12}+\sigma_{12}\sigma_{22} \hspace{20pt} a_3 = \sigma_{12}^2+\sigma_{22}^2,
\end{align*}
we obtain 
\begin{widetext}
\begin{align*} 
	\hat{R}_{\xi\xi}(\tau)&=\left((a_1+a_3)\upsilon ^2 +2 a_2 \gamma  \upsilon+2 a_1 \gamma ^2 \right)\cos (\upsilon\tau)+   \left(  (a_3-a_1)\gamma\upsilon+2 a_2 \gamma^2 \right)\sin (\upsilon\tau) \\  
	\hat{R}_{\xi\zeta}(\tau)&=\left( (a_3-a_1)\gamma\upsilon +2 a_2 \gamma^2 \right)\cos (\upsilon\tau)- \left( (a_1+a_3)\upsilon ^2+2 a_2 \gamma  \upsilon+2 a_1 \gamma^2 \right) \sin (\upsilon\tau)\\
	\hat{R}_{\zeta\xi}(\tau)&=\left( (a_3-a_1)\gamma \upsilon +2 a_2 \gamma^2 \right)\cos (\upsilon\tau) + \left( (a_1+a_3)\upsilon ^2-2 a_2 \gamma  \upsilon +2 a_3 \gamma ^2\right)\sin (\upsilon\tau)\\
	\hat{R}_{\zeta\zeta}(\tau)&=\left( (a_1+a_3)\upsilon ^2-2 a_2 \gamma  \upsilon +2 a_3 \gamma ^2\right)\cos (\upsilon\tau)+   \left( (a_1-a_3)\gamma\upsilon -2 a_2 \gamma^2 \right)\sin (\upsilon\tau).
\end{align*}
\end{widetext}



%

\end{document}